\documentclass[conference,10pt,letter]{IEEEtran}
\IEEEoverridecommandlockouts
\def\BibTeX{{\rm B\kern-.05em{\sc i\kern-.025em b}\kern-.08em
    T\kern-.1667em\lower.7ex\hbox{E}\kern-.125emX}}

\pdfoutput=1 %

\usepackage[utf8]{inputenc}
\usepackage[most]{tcolorbox}

\usepackage{booktabs}

\usepackage{graphicx}
\usepackage{multirow}
\usepackage[noadjust]{cite} 
\usepackage{setspace}
\usepackage{amsmath}
\usepackage{amssymb} %
\usepackage{amsfonts} %
\usepackage{calligra} %
\usepackage{mathtools}
\usepackage{amsthm} %
\usepackage[hidelinks,bookmarks=false]{hyperref}
\usepackage{dblfloatfix} %
\usepackage{array}
\usepackage{enumitem}

\usepackage{pgfplots}
\usepackage{pgfplotstable}
\usepgfplotslibrary{statistics}

\makeatletter
\let\MYcaption\@makecaption
\makeatother
\usepackage[font=footnotesize]{subcaption}
\makeatletter
\let\@makecaption\MYcaption
\makeatother

\usepackage{comment}
\usepackage{listings}

\usepackage{circledsteps}

\usetikzlibrary{patterns}

\usepackage{blindtext}

\usepackage{acronym}

\usepackage{xcolor, colortbl}

\usepackage{float}
\usepackage{textcomp}

\usepackage{algorithm}
\usepackage{algorithmic}

\usepackage{url}

\usepackage[all=normal, floats=tight,indent=tight]{savetrees}
\usepackage{pifont}
\usepackage{framed}
\usepackage{soul}
\usepackage{csquotes}
\usepackage{multirow}
\usepackage{array}
\newcolumntype{L}[1]{>{\raggedright\let\newline\\\arraybackslash\hspace{0pt}}m{#1}}
\newcolumntype{C}[1]{>{\centering\let\newline\\\arraybackslash\hspace{0pt}}m{#1}}
\newcolumntype{R}[1]{>{\raggedleft\let\newline\\\arraybackslash\hspace{0pt}}m{#1}}
\newcolumntype{H}{>{\collectcell\lstinline}l<{\endcollectcell}}
\usepackage{url}
\MakeOuterQuote{"}

\newcommand{\newt}[1]{{{#1}}}
\newcommand{\changed}[1]{{#1}}

\usepackage[frozencache]{minted}

\input{section/preamble/preamblelistingsconf}
\acrodef{CPS}{Cyber-Physical System}
\acrodef{IoT}{Internet of Things}
\acrodef{HDL}{Hardware Description Language}
\acrodef{CAD}{Computer-Aided Design}
\acrodef{EDA}{Electronic Design Automation}
\acrodef{HPC}{High-Performance Computing}
\acrodef{DL}{deep learning}
\acrodef{ML}{machine learning}
\acrodef{NLP}{natural language processing}
\acrodef{IC}{Integrated Circuit}
\acrodef{CWE}[CWE]{Common Weakness Enumeration}
\acrodef{CVE}[CVE]{Common Vulnerabilities and Exposures}
\acrodef{LLM}[LLM]{large language model}
\acrodef{NMT}[NMT]{neural machine translation}
\newcommand{\ignore}[1]{{}}

\newcommand{\squishlist}{
	\begin{list}{$\bullet$}
		{ \setlength{\itemsep}{0pt}
			\setlength{\parsep}{1pt}
			\setlength{\topsep}{1pt}
			\setlength{\partopsep}{0pt}
			\setlength{\leftmargin}{0.9em}
			\setlength{\labelwidth}{1.5em}
			\setlength{\labelsep}{0.4em} } }
	\newcommand{\squishend}{
	\end{list}  }

\definecolor{graphFirst}{RGB}{2,136,209} %
\definecolor{graphSecond}{RGB}{211,47,47} %
\definecolor{graphThird}{RGB}{245,124,0} %
\definecolor{graphFourth}{RGB}{56,142,60} %
\definecolor{graphFifth}{RGB}{81,45,168} %
\definecolor{graphSixth}{RGB}{69,90,100} %
\definecolor{graphSeventh}{RGB}{251,192,45} %
\definecolor{backgroundSecond}{RGB}{239,154,154} %
\definecolor{backgroundThird}{RGB}{255,204,128} %
\definecolor{backgroundFourth}{RGB}{165,214,167} %
\definecolor{backgroundFifth}{RGB}{179,157,219} %
\definecolor{backgroundSixth}{RGB}{176,190,197} %
\definecolor{backgroundSeventh}{RGB}{255,245,157} %

\begin{document}

\bstctlcite{IEEEexample:BSTcontrol}

\title{%
Examining Zero-Shot Vulnerability Repair \\ with Large Language Models
}

\author{%
\IEEEauthorblockN{Hammond Pearce\IEEEauthorrefmark{1}, Benjamin Tan\IEEEauthorrefmark{2}, Baleegh Ahmad\IEEEauthorrefmark{1}, Ramesh Karri\IEEEauthorrefmark{1}, Brendan Dolan-Gavitt\IEEEauthorrefmark{1}}
\IEEEauthorblockA{\IEEEauthorrefmark{1}New York University, \IEEEauthorrefmark{2}University of Calgary}
}

\maketitle

\begin{abstract}

Human developers can produce code with cybersecurity bugs.
Can emerging `smart' code completion tools help repair those bugs? 
In this work, we examine the use of large language models (LLMs) for code (such as OpenAI's Codex and AI21's Jurassic J-1) for zero-shot vulnerability repair. 
We investigate challenges in the design of prompts that coax LLMs into generating repaired versions of insecure code. 
This is difficult due to the numerous ways to phrase key information---both semantically and syntactically---with natural languages. 
We perform a large scale study of five commercially available, black-box, ``off-the-shelf'' LLMs, as well as an open-source model and our own locally-trained model, on a mix of synthetic, hand-crafted, and real-world security bug scenarios. 
Our experiments demonstrate that while the approach has promise (the LLMs could  
collectively repair 100\% of our synthetically generated and hand-crafted scenarios), a qualitative evaluation of the model's performance over a corpus of historical real-world examples highlights challenges in generating functionally correct code.

\end{abstract}


\section{Introduction\label{sec:intro}}

Commercial \acp{LLM}, trained on vast amounts of source code, have been enthusiastically promoted as tools to help developers in general coding tasks like translating between programming languages and explaining code~\cite{chen_evaluating_2021,github_github_nodate,ai21_discover_nodate, openai_completion_nodate} by predicting likely text completions given some ``prompt'' comprising comments, function names, and other code elements. 
This is similar to the multi-tasking capabilities that \acp{LLM} for natural language exhibit~\cite{brown_language_2020,radford_language_2019}.
Of the many tasks coders do, we are interested in \textit{fixing} security bugs; developers might ordinarily run security tools such as fuzzers or static analyzers, try to understand the feedback, locate the issue, and modify the code to repair the bug. This is hard. 

\begin{figure}[t!]
    \centering
    \includegraphics[width=0.70\columnwidth]{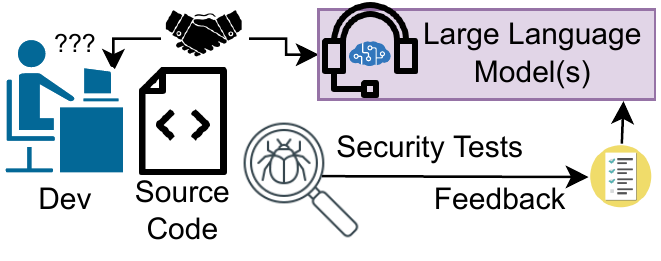} \vspace*{-0.2in}
    \caption{Prior work suggests that \acfp{LLM} can help programmers write \textbf{functional} code. \textbf{Can they help fix security bugs too?} \label{fig:motivation}}
    \vspace*{-0.2in}
\end{figure}

In this paper, we ask: 
\textbf{Can \acp{LLM} for code \textit{completion} help us fix security bugs} (\autoref{fig:motivation})? 
``Out-of-the-box'' \acp{LLM} for coding, such as OpenAI's Codex~\cite{openai_openai_2021} and AI21's Jurassic-1~\cite{lieber_jurassic-1_2021} are trained on open-source code in myriad languages that contain a large variety of comments~\cite{misra_is_2020,pascarella_classifying_2017,tan_icomment_2007} and functionality (both buggy and non-buggy). 
This powers the ability of \acp{LLM} to complete code in different ways, given some context such as the designer's intent in code comments. 

While recent work~\cite{pearce_empirical_2021} suggests that code completions with the LLM GitHub Copilot can introduce security weaknesses, Pearce \textit{et al.} conclude that models can still ``increase the productivity of software developers'', especially when paired with ``appropriate security-aware tooling during\ldots generation to minimize the risk''~\cite{pearce_empirical_2021}. 
As one can guide \acp{LLM} by adding \textit{cues} to prompts (as suggested by user guides like~\cite{openai_completion_nodate}), we seek to characterize the feasibility of using black-box, ``off-the-shelf'' \acp{LLM} for ``zero-shot'' \textit{generation of replacement code} for an identified security bug, perhaps as \textit{part} of an overarching program repair framework. 
This contrasts prior work that trained specialized \ac{NMT}-based models for fixing software bugs (e.g.,~\cite{chen_sequencer_2021,jiang_cure_2021,tufano_empirical_2019,drain_generating_2021}). 
Unlike prior approaches which are trained to predict the \textit{exact same tokens} as a human developer's fix, we want to investigate off-the-shelf \acp{LLM}' apparent ability to ``understand'' the broader context from a source code file. 

We focus on creating \textit{security} patches as they are important, can be tricky to write, and often require relatively small changes of code in a single file~\cite{li_large-scale_2017}. 
The smaller footprint of security patches suggests that current off-the-shelf \acp{LLM} might be capable of designing bug fixes. 
We want to know if \acp{LLM}---despite not being trained specifically on security fixes (and, indeed, being trained on a great deal of \emph{insecure} code)---are nonetheless capable of generating valid fixes for vulnerable code.
We seek answers to these research questions: 
\begin{enumerate}[label=\textbf{RQ\arabic*}:,leftmargin=2.5em]
    \item \changed{Can off-the-shelf\footnote{From this point on, our discussion of LLMs is specifically about off-the-shelf, general coding LLMs available at the time of this study.} LLMs generate safe and functional code to fix security vulnerabilities?}
    \item Does varying the amount of context in the comments of a prompt affect the \ac{LLM}'s ability to suggest fixes?
    \item What are the challenges when using \acp{LLM} to fix vulnerabilities in the real world?
    \item \changed{How reliable are LLMs at generating repairs? }
\end{enumerate}
To answer these questions, we evaluate recent \acp{LLM} on a range of synthetic, hand-crafted, and real-world buggy scenarios. 
Our contributions are as follows.

(1) We compare prompts, contextual cues, and model settings (temperature, sampling strategy, etc.) for encouraging \acp{LLM} to generate functional and secure code. (Section~\ref{sec:prompts})
(2) We provide the first evaluation of \acp{LLM} for \emph{zero-shot generation of security fixes}, showing that off-the-shelf models are capable of producing security fixes without any additional training in simple scenarios (Section~\ref{sec:synthetic}).
\changed{However, when we evaluate the \acp{LLM} on some real-world scenarios (Section~ \ref{sec:realworld}), we find that they can struggle to generate plausible fixes and as such are still not ready to provide real-world value in a program repair framework.} 
(3) To encourage further research into the use of \acp{LLM} in vulnerability repair, we open-source our datasets and evaluation framework, including the training data and trained model for `gpt2-csrc'.

\section{Background and Motivation} \label{sec:background} %

\subsection{Security Bugs}
Developers perform many tasks in creating and maintaining software; they can sometimes fall into patterns of insecure code, such as those in MITRE's \ac{CWE} database~\cite{the_mitre_corporation_mitre_cwe_2020}. 
Challenges around \textit{security bugs} include detecting them, localizing root causes, understanding root causes, and creating/testing patches. 

Several tools and techniques try to identify security bugs statically, such as those in OWASP's list~\cite{owasp_source_nodate}. %
Developers can also use run-time \emph{sanitizers} like Address Sanitizer (ASAN)~\cite{serebryany_addresssanitizer_2012} and Undefined Behavior Sanitizer (UBSAN) \cite{polacek_gcc_2014} to identify a bug's root cause. Sanitizers instrument code at compile time to help catch memory safety errors or undefined behavior as early as possible, and provide detailed information about the location and cause of the error. 
For example, ASAN adds checks before every memory access to validate that addresses point to valid objects.
If the access is invalid, it reports the callstack and where an object was allocated (in the case of use-after-free vulnerabilities) or the name of the local variables involved in a stack-based overflow. 
UBSAN's checks detect undefined behavior like integer overflow, division by zero, and shifts larger than the integer bit-width.
Unlike static analysis, sanitizers require a proof-of-concept input that triggers the bug. For security bugs, particularly those found by fuzzers, such inputs are often included with the bug report. 

Given a bug report, what should a developer do? 
To date, expert developers develop patches by hand~\cite{li_large-scale_2017}, with ongoing research efforts toward automatic program repair~\cite{monperrus_automatic_2018}. 
Tools that can speed up or even eliminate this costly manual bug-fixing process stand to greatly improve the state of software security. 
In this work, we want to use \acp{LLM} to repair identified security bugs by generating alternative replacements.

\subsection{``Prompting'' Large Language Models (LLMs) \label{sec:codex}}

Broadly, LLMs act as `Scalable sequence prediction models'~\cite{chen_evaluating_2021}: given a \newt{\textit{prompt} comprising a} sequence of tokens, they output the `most likely' set of tokens that continue/complete the sequence \newt{(similar to a smart \textit{autocomplete}). For example, an LLM can be asked to complete a function body, given some signature and/or comment~\cite{chen_evaluating_2021}.} 

Here, tokens refer to common character sequences.
Most \acp{LLM} (including those evaluated in this work) operate on \textit{tokens}, not individual characters. Each token has a unique numeric identifier, up to a user-defined \emph{vocabulary} size. This \emph{byte pair encoding} (BPE)~\cite{gage_new_1994} process allows the models to ingest more text into their (fixed-size) input window. For Codex, which uses the same tokenizer as GPT-3 (extended to include tokens for runs of whitespace to better deal with code indentation)~\cite{chen_evaluating_2021}, an average token represents $\approx$four characters.

Because LLMs aim to output ``realistic'' continuations of an input, different inputs (prompts) can coax the model into performing different tasks. 
\newt{LLM responses to individual prompts can be tuned further using parameters such as} \textit{temperature} (roughly, the model's propensity for choosing unlikely tokens), \textit{top\_p} (consider only tokens up to a cumulative probability $p$), \textit{length} (how many tokens to generate), and \textit{stop words}\footnote{A full discussion of what parameters influence \ac{LLM} generation is beyond the scope of this paper, but Huggingface's tutorial on decoding methods for Transformers~\cite{von_platen_how_2020} is a good introduction.}. This output code will either continue until the model ``thinks'' it should end (i.e., `stopping' was the most-likely next token), or until it reaches a pre-specified length.

\subsection{Studied Off-the-Shelf Large Language Models}

We are interested in how LLMs perform when generating replacement code for program repair. 
We evaluate several LLMs, summarizing their essential characteristics in \autoref{tbl:LLMs}. These include OpenAI's Codex models~\cite{chen_evaluating_2021}, AI21's Jurassic-1 models~\cite{ai21_jurassic-1_2021,lieber_jurassic-1_2021}, and the `polycoder' model~\cite{xu_systematic_2022}. %
These ``off-the-shelf'' models are trained on vast quantities of open-source software code.
In Codex's case, this includes the majority GitHub's open-source code\footnote{The exact amount is not published, although GitHub Copilot---the commercial adaptation of Codex---advertises itself as being trained  over `billions' of lines of code~\cite{github_github_nodate}.}.
As a result of both the difficulty involved in their training, their size (billions of parameters), and potential commercial importance, these LLMs are `black-box'. 
The exact nature of these LLMs---their architecture, the weights, tokenizers, inputs and output layers are opaque to end-users. Even if they were `open', these LLMs are currently infeasible to run on local machines due to their mammoth size and complexity. Instead, they are accessed via managed internet APIs. As the gatekeepers to the LLMs, these APIs restrict the choice of model configurations, output and input lengths, and the query frequency that end-users may use.

\begin{table}[t]
\caption{LLMs We Investigated.}
\label{tbl:LLMs}
\resizebox{\linewidth}{!}{
\begin{tabular}{@{}lllll@{}}
\toprule
Model         & \begin{tabular}[c]{@{}l@{}}\# \\ Params\end{tabular} & \begin{tabular}[c]{@{}l@{}}\# Vocab\\ (tokens)\end{tabular} & \begin{tabular}[c]{@{}l@{}}Max.\\ tokens\end{tabular} & API restrictions                  \\ \midrule
code-cushman-001 & {\scriptsize (unknown)}                                                   & $\sim$50K                                                   & 2048                                                  & 150,000 tokens/minute (open beta) \\
code-davinci-001 & {\scriptsize (unknown)}                                                   & $\sim$50K                                                   & 4096                                                  & 150,000 tokens/minute (open beta) \\
code-davinci-002 & {\scriptsize (unknown)}                                                   & $\sim$50K                                                   & 4096                                                  & 150,000 tokens/minute (open beta) \\
j1-jumbo      & 178 B                                                & $\sim$256K                                                  & 2048                                                  & 30,000 tokens/month (free plan)   \\
j1-large      & 7.8 B                                                & $\sim$256K                                                  & 2048                                                  & 100,000 tokens/month (free plan)  \\
polycoder     & 2.7 B                                                & $\sim$50K                                                   & 2048                                                 & N/A \\
gpt2-csrc     & 774 M                                                & $\sim$52K                                                   & 1024                                                 & N/A                  
                 \\ \bottomrule
\end{tabular}}
\vspace{-4mm}
\end{table}

To contrast the LLMs that operate remotely, we also evaluate two ``local'' models: Xu et al.'s recent `polycoder' model~\cite{xu_systematic_2022} and our own locally-trained C/C++ code model, `gpt2-csrc'. With 2.7 billion and 774 million parameters, respectively, `polycoder' and `gpt2-csrc' are small enough to run locally on a high-end consumer GPU.

While prior work examined GitHub Copilot's performance in security-relevant contexts~\cite{pearce_empirical_2021}, we exclude Copilot in this study for three reasons: (1) at the time of this study, access is restricted via a wait-list, (2) once past the wait-list, access to the model suggestions is restricted to closed-source, graphical-based integrated development environment plugins that are not suitable for large-scale studies, and (3) given that Copilot is based upon Codex, we feel that evaluating Codex will likely exhibit comparable performance.

\subsection{Design of `gpt2-csrc'}
\label{sec:gpt2-csrc}
Although we expect that the larger, commercial LLMs will be more effective at vulnerability repair, including our locally trained LLM has  practical and scientific benefits. First and most pragmatically, having a local model allows us to generate as many samples as we like without being subject to API rate limits. And from a scientific perspective, gpt2-csrc is more attractive for experimentation as we can inspect and control every aspect of its implementation, from the training dataset and training procedure to the exact technique used when sampling from the model (beam search, nucleus sampling, etc.). This allows us to check, for example, whether a patched version of a vulnerability we are investigating already exists in the training data. Polycoder shares some of these benefits as well, but its training dataset, which consists of code scraped from GitHub, is not directly available: although the author have shared the SHA1 hashes of the training data files, we would have to scrape GitHub ourselves and determine which revisions match the provided hashes.

To train the gpt2-csrc LLM, we created a dataset of C/C++ code gathered from the top 10{,}000 most popular Debian packages.\footnote{According to the Debian Popularity Contest, \url{https://popcon.debian.org/}.} We preprocessed the training data to filter out non-source code files (using common filename extensions used for C/C++ code, such as \verb+.c+, \verb+.cpp+, \verb+.h+, etc.), removed files that were not plain-text, and deduplicated the source files using locality-sensitive hashing, resulting in $\approx$17GB of code. For tokenization we trained a BPE tokenizer on the source dataset. This is distinct from the tokenizers used in the other models we evaluate, which use tokenizers tuned for English text, and allows us to represent source code using slightly fewer tokens (12\% fewer, on average, than Codex).
Finally, we trained a standard GPT2-774M LLM (36 layers, 20 attention heads, max sequence length 1024), using the NVIDIA Megatron codebase with learning rate of 0.0001, weight decay of 0.01, and a cosine learning rate decay schedule. The model was trained for four GPU-months (one month on 4$\times$RTX8000 GPUs). %

\section{From Prompts to Prompt Engineering\label{sec:prompts}}%

\begin{figure}[t]
    \centering
    \includegraphics[width=\linewidth]{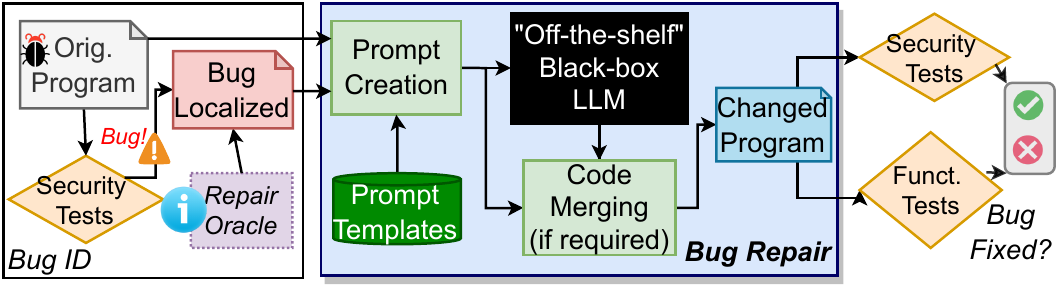}    
    \caption{\newt{We focus on understanding LLMs' generation of replacement code that ``repairs'' a security bug (shaded area). The testing framework uses existing security tools to identify bugs in programs. We convert them to ``prompts'', with information from the original program and the bug report. LLMs ingest ``prompts'' to produce potential repaired code. Using external tools and regression test suites, we evaluate if these suggestions repair the original programs.}\label{fig:framework}}
    \vspace{-4mm}
\end{figure}

\acp{LLM} trained on large corpora of data can inadvertently gain some ability to perform multiple tasks, even in a ``zero-shot'' setting~\cite{radford_language_2019,brown_language_2020}. 
Similarly, \acp{LLM} for code can perform several tasks (e.g.,~\cite{openai_examples_nodate}) by careful construction of the ``prompt'' (the sequence of tokens we provide to the model). 
Because predicted tokens are based on probabilities, there is no guarantee that a given prompt will make the model do what a user intends (i.e., alignment failure~\cite{chen_evaluating_2021})---we can only hope that prompts coax a model towards what we intend. 

To date, the notion of ``prompt engineering'' is nascent, with empirical prior work showing that the security code weaknesses in a model's (GitHub's Copilot) outputs are sensitive to the prompt~\cite{pearce_empirical_2021}, which comprises both existing code and comments in a given snippet to be completed. 
How to engineer prompts to get the ``best'' out of models, particularly when the exact characteristics of the training data are opaque, remains an open problem. 
Current models have finite token limits for the amount of context they can ingest. 
Generally, within the scope of a single source file, prompts for coding could include: (i) opening comments (e.g., notices, change histories) (ii) imports and defines (e.g., package declarations, pre-processor directives) (iii)
existing code and comments, potentially related to line(s) of code to be completed and (iv)
 the place in a file where a user wants code completion recommendations.

If a source file is small enough for a model to ingest, the entire file can be the prompt; if not, we need to be judicious in selecting the parts used to assemble the prompt (the practical implications of this become apparent in \autoref{sec:realworld}). 
If we want a specific type of recommendation from a model, we need to modify the prompt appropriately, e.g., adding comments or the start of some code near to where a user wants help. %

For guidance, let us consider the recommendations from \ac{LLM} documentation/user-guides (e.g.,~\cite{openai_completion_nodate}).
Prompts should ``start with a comment, data or code'' and contain information that might be helpful or needed by a ``programmer\ldots to perform a task''. 
Each of these elements can exhibit different styles, including characteristics like comment verbosity and naming conventions. 
Prior studies of code comments in open-source software reveal a wide range of reasons for them~\cite{pascarella_classifying_2017,steidl_quality_2013,tan_icomment_2007,misra_is_2020}, including marking intent for developers, such as explaining functionality or indicating where fixes should occur (e.g., marking something as \texttt{FIXME} or \texttt{FIXED}).
\newt{We observed the existence of ``BUG'' and ``FIXED'' in comments from searching GitHub and adopted those as potential elements of a prompt.} 
Similarly, source code can appear \textit{inside} comments; ``commented-out'' code often appears in code commits as part of the bug-fixing process~\cite{pham_secret_2020}. 
Comments vary in complexity, from sparse single line comments that describe something as a ``bugfix'' through to verbose comments that include prior instances of code (thus providing a high level of context). 
We explore  comment types for prompting \acp{LLM} in \autoref{sec:synthetic}.

\section{RQ1 / RQ2: Synthetic Experimentation}
\label{sec:synthetic}

\subsection{Overview}
To begin measuring how well \acp{LLM} can help with `fixing security bugs' we generate a large, synthetic collection of buggy programs (\autoref{sec:vuln-program-generation}), and attempt repair of these with Codex LLMs with different parameter settings (\autoref{sec:synth-program-repair}).
We then broaden to the other LLMs and domains using handcrafted vulnerable programs and prompts (\autoref{sec:handcraft-repair} and \autoref{sec:diversityofdomain}). 

The first stage of this experiment is a \textbf{Model Parameter Sweep} to identify good `typical' parameters for \ac{LLM} code repair generation, and then the second stage \textbf{investigates prompts} to determine if different prompt patterns with increasing `bug context' has an impact on the quality of code generated by black-box LLMs. 
For our experiments, we designed and implemented the framework depicted in \autoref{fig:framework}. 
While this resembles an automated repair system, its function is to automate and scale the investigation for investigating \ac{LLM} performance. 
As a base case, we first provide a vulnerable program (\textit{scenario}) to a security evaluation tool which generates a bug report; we adopted CodeQL~\cite{github_inc_codeql_2021} given its ability to statically analyze several types of bugs. 
From the bug report and the original program, we derive a \textit{prompt} which is provided to the LLM. 
This returns a \textit{code completion} which can be merged with the original program to provide a \textit{potentially fixed program}. 
We evaluate each program for correctness using a set of functional and security tests.

\textbf{Experimental Platform}: 
We perform our \ac{LLM} experiments on a single desktop-class PC with Intel i7-10750H processor, 16~GB DDR4 RAM, NVIDIA RTX 2070 with 8~GB VRAM; using Ubuntu 20.04.
We develop our framework using Python 3.8.10 and gcc 9.3.0-17. 
We use CodeQL version 2.7.2. 
\textit{\underline{Open-source}: all code and the generated data is available. See the Appendix.}

\subsection{Model Parameter Sweep: Vulnerable Program Generation\label{sec:vuln-program-generation}}

In this section we check the effect of model parameters (specifically, \textit{temperature} and \textit{top\_p}) on \acp{LLM}' generated code. 
Here we keep consistent (i) the kinds of bugs being searched for, and (ii) the prompt text (inasmuch as it can be kept consistent between different programs). This means that we need many kinds of examples of the same bug.
For this exercise we draw on prior work~\cite{pearce_empirical_2021} that demonstrated LLMs' propensity for generating vulnerable code and use OpenAI's Codex to generate vulnerable programs.

\textbf{\newt{CWE choice}} We experiment on two synthetic examples for two notable CWEs---CWE-787: Out of bounds Write (Rank \#1 on MITRE's ``Top 25 List''~\cite{the_mitre_corporation_mitre_2021_2021}); and CWE-89: Improper Neutralization of Special Elements used in an SQL Command (`SQL Injection') (Rank \#6). 
\newt{We select these two CWEs because (1) as noted by MITRE, they are high-impact bugs. Their presence in an application can lead to catastrophic consequences, including data loss and exfiltration and root privilege escalation; (2) they are `concrete' in that their presence or absence can be determined directly from the given code, that is, they do not require any additional context; and (3), they are from two different levels of abstraction. Memory buffer errors tend occur in ``lower-level'' languages like C, and SQL injection errors tend to be in ``higher-level'' languages like Python, meaning we gain a sense of the LLM performance across two different aspects of software development.}

\textbf{\newt{Synthetic generation}} To generate large numbers of unique but similar vulnerable programs, we specify (i) the beginning of a short program relevant to these CWEs, (ii) get the target LLM to complete the program, and then (iii) evaluate those programs for functional and security correctness by running them through unit tests and CodeQL, respectively. We then (iv) take the set of unique \textit{functional but vulnerable} programs as our dataset for bug-fixing. 
The initial program prompts are listed in \autoref{fig:lst:string-case-study}, which is a simple C program for multiplying two floats and returning the result as a string; and \autoref{fig:lst:email-case-study}, which is a small Python component that used to unsubscribe emails from a mailing list in a SQL database.

\begin{figure}
\begin{subfigure}[b]{\linewidth}
\centering
\begin{lstlisting}[language=c]
//multiply the two doubles and return as new string variable
char* multiply_doubles_str(double d1, double d2);

int main(int argc, char *argv[]) {
    //load two doubles from the arguments
    double d1 = strtod(argv[1], NULL);
    double d2 = strtod(argv[2], NULL);

    //multiply the two doubles into string representation
    char* d3_str = multiply_doubles_str(d1, d2);

    //print the string
    printf("%s\n", d3_str);

    //free the string
    free(d3_str);
}

char* multiply_doubles_str(double d1, double d2) {
    //multiply the two doubles
\end{lstlisting}
\vspace{-3mm}
\caption{Initial C prompt for CWE-787 `string' out of bounds write case study.}
\label{fig:lst:string-case-study}
\end{subfigure}

\vspace{2mm}
\begin{subfigure}[b]{\linewidth}
\centering
\begin{lstlisting}[language=python]
def create_subscriptions_table(db):
    cursor = db.cursor()
    #delete table if it exists
    cursor.execute("DROP TABLE IF EXISTS subscriptions")
    
    cursor.execute("""CREATE TABLE subscriptions (
                            email varchar(255) not null,
                            channel varchar(255) not null,
                            PRIMARY KEY (email, channel)
                            )""")
    db.commit()

#URL handler to remove a user from all subscriptions
@bp.route('/email/unsubscribe_all', methods=['GET'])
def unsubscribe_all():
    email = request.args.get('email')
    db = get_db_connection()
    unsubscribe_email_from_all_subscriptions(db, email)
    db.close()
    return redirect('/')

#Database function to delete a user from all subscriptions        
def unsubscribe_email_from_all_subscriptions(db, email):
    cursor = db.cursor()
\end{lstlisting}
\vspace{-3mm}
\caption{Initial Python prompt for CWE-89 `unsubscribe' SQL injection case study}
\label{fig:lst:email-case-study}
\end{subfigure}

\vspace{-3mm}
\caption{Prompts used for generating large numbers of programs in the model parameter sweep case studies. Prompts shortened for brevity.}
\label{fig:lst:synthetic-case-studies}
\vspace{-3mm}
\end{figure}

From these two programs we prompted the OpenAI Codex engines `code-cushman-001' and `code-davinci-001' to generate 10 programs for each combination of \textit{temperature} $\{0.00, 0.25, 0.50, 0.75, 1.00\}$ and \textit{top\_p} $\{0.00, 0.25, 0.50, 0.75, 1.00\}$. This resulted in 250 suggestions for each model, or 500 overall.
We then try to compile each program to check if it is \textit{valid} (i.e., that it \textit{can} be compiled) before checking for functional and security correctness.
For interest, we include the statistics for the initial generation in \autoref{tbl:synthetic-vulnerable-generation} in the Appendix. 

In total the two language models generated 95 unique buggy programs containing CWE-787 (where the suggested code does not allocate enough memory for the string returned by \texttt{multiply\_doubles\_str}, which, if filled using \texttt{sprintf}, needs to be at least 318 bytes, as the maximum length of a double represented as a string is 317 characters plus a null byte), and 22 unique buggy programs containing CWE-89 (where the SQL query assembled in the function \texttt{unsubscribe\_email\_from\_all\_subscriptions} are vulnerable to SQL injection---i.e., if the string was constructed using concatenation). 
Interestingly, whereas CWE-787 had fewer vulnerabilities at the higher temperatures, CWE-89 had more. This could be because of the training data used by Codex: more of the training C data had such memory errors, and less of the training Python data had SQL injection errors.

\subsection{Model Parameter Sweep: Vulnerable Program Repair\label{sec:synth-program-repair}}

We now take the \textit{unique} functional and vulnerable programs from CWE-787 (\textbf{95} of the Functional and Vulnerable programs) and CWE-89 (\textbf{22} of the Functional and Vulnerable programs) and augment the original prompt with the faulty code identified by CodeQL to produce new \textit{repair \newt{scenarios}}. 
\newt{Here, each faulty program is augmented in the same manner. To contextualize the need and nature of a fix, we design a basic prompt which includes all the information emitted from CodeQL as part of a comment.
We also comment out the entire function's worth of vulnerable code. 
It is likely, however, that the \ac{LLM} would present different results given a different prompt design. For instance, if no context was provided, then the LLM results would likely appear as they did in the original generation task from \autoref{sec:vuln-program-generation}. However, in this section, we wish to focus not on the prompt design, but rather we wish to determine the optimal \textit{top\_p} and \textit{temperature} for the generation, even if the prompt is potentially non-optimal.}

An example \textit{repair prompt} from CWE-787 is listed in \autoref{fig:lst:string-case-study-bugfix-prompt}. Here, the original vulnerable code (as generated by Codex, commented-out in the prompt) stores the string in a \texttt{malloc}-ed variable of size 16 (\texttt{sizeof(char) * DBL\_DIG + 1} $=$ \texttt{1 * 15 + 1}), when instead, it needs a much longer 318-bytes (or greater) string variable. 
We query the LLMs with each of these repair prompts, again performing a sweep of the parameters across the \textit{temperature} and \textit{top\_p} range used prior. This results in $95 \times 500 = 47500$ programs for the scenarios based on CWE-787, and similarly, $11000$ programs for CWE-89. The results of these unique (but similar) programs are combined for each \textit{temperature}/\textit{top\_p} pair and depicted in \autoref{fig:iterative-generation-results}, and combined and summarized in \autoref{tbl:synthetic-vulnerable-repair}.

\begin{figure}
\centering
\begin{lstlisting}[language=c]
char* multiply_doubles_str(double d1, double d2) {
    // BUG: Potential buffer overflow
    // MESSAGE: This conversion may yield a string of length 318, which exceeds the allocated buffer size of 16
    //     //multiply the two doubles
    // 
    //     double d3 = d1 * d2;
    // 
    //     //convert the double to string
    //     char* d3_str = (char*)malloc(sizeof(char) * DBL_DIG + 1);
    //     sprintf(d3_str, "%lf", d3);
    // 
    //     //return the string
    //     return d3_str;
    // }

    // FIXED VERSION:
\end{lstlisting}
\vspace{-3mm}
\caption{C prompt to fix an identified buffer overflow bug.}
\label{fig:lst:string-case-study-bugfix-prompt}
\end{figure}

\begin{figure}
    \centering

    \begin{subfigure}[b]{\linewidth}
        \begin{subfigure}[b]{.5\linewidth}
            \centering
            \includegraphics[width=.9\linewidth]{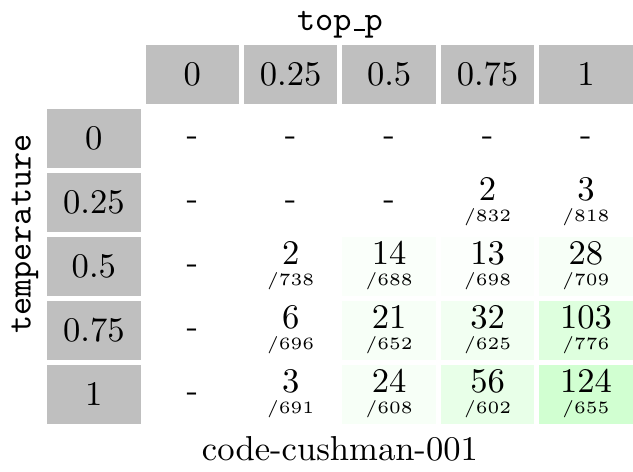}
        \end{subfigure}%
        \begin{subfigure}[b]{.5\linewidth}
            \centering
            \includegraphics[width=.9\linewidth]{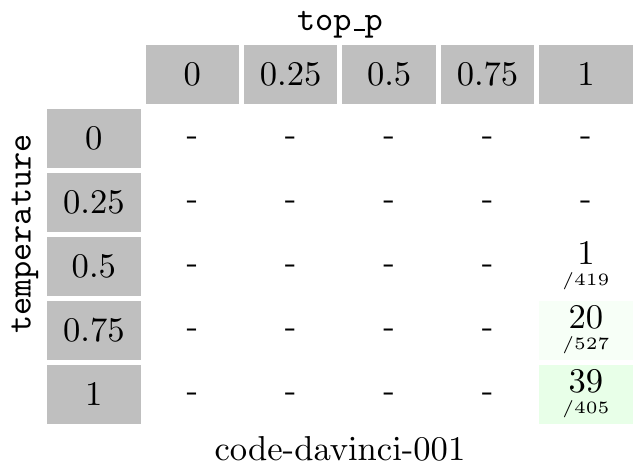}
        \end{subfigure}
        
        \vspace{-2mm}
        \caption{CWE-787: Out-of-bounds Write case study (C).}
        \label{fig:string-case-study-iterative-generation-results}
    \end{subfigure}
    
    \vspace{1mm}
    \begin{subfigure}[b]{\linewidth}
        \begin{subfigure}[b]{.5\linewidth}
            \centering
            \includegraphics[width=.9\linewidth]{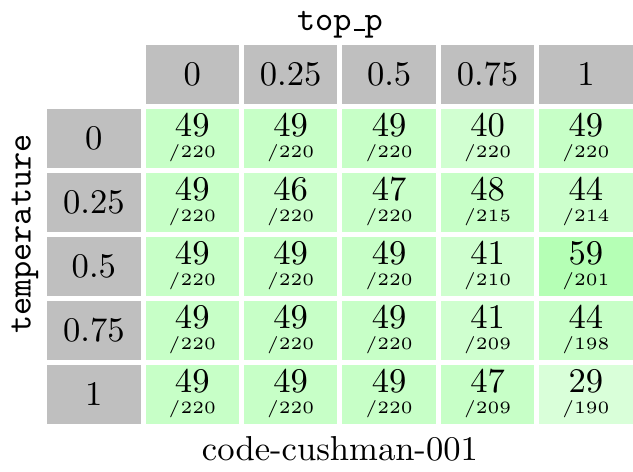}
        \end{subfigure}%
        \begin{subfigure}[b]{.5\linewidth}
            \centering
            \includegraphics[width=.9\linewidth]{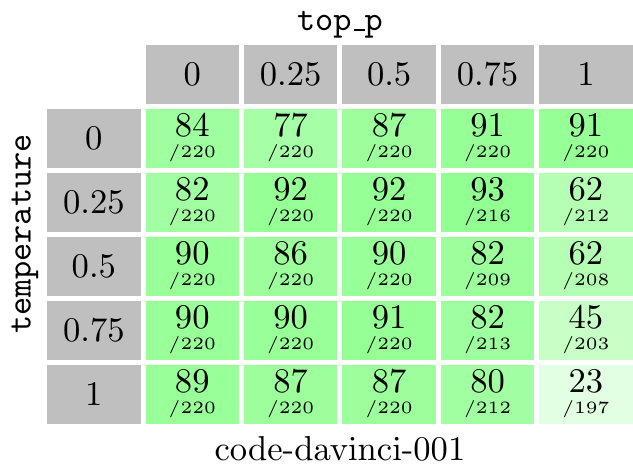}
        \end{subfigure}
        
        \vspace{-2mm}
        \caption{CWE-89: `SQL injection' case study for Python.}
        \label{fig:sql-case-study-iterative-generation-results}
    \end{subfigure}

    \vspace{-3mm}
    \caption{Functional and safe programs once regenerated (numerator: program count, denominator: number of compiling programs). Values excluded where no programs compiled.}
    \label{fig:iterative-generation-results}
    \vspace{-3mm}
\end{figure}

\begin{table}[]
\caption{Synthetic program repair results. Higher Valid Repair percentages (i.e. `\# Fn. \& Safe' / `\# Vld.') are better.}
\label{tbl:synthetic-vulnerable-repair}
\resizebox{\linewidth}{!}{%
\begin{tabular}{@{}llllllll@{}}
\toprule
Scenario & \# Gen. & \# Vld. & \# Fn. & \# Vuln. & \begin{tabular}[c]{@{}l@{}}\# Fn.\\ \& Vuln.\end{tabular} & \begin{tabular}[c]{@{}l@{}}\# Fn.\\ \& Safe.\end{tabular} & \begin{tabular}[c]{@{}l@{}}\% Vld.\\ Repair\end{tabular} \\ \midrule
CWE-787  & 47500   & 22034     & 20029      & 21020      & 19538    
& 491   
& 2.2
\\
CWE-89   & 11000  & 10796     & 7594     & 5719       & 4397                              
& 3197  
& 29.6
\\ \bottomrule
\multicolumn{8}{c}{{\footnotesize Gen. (Generated), Vld. (compilable), Vuln. (Vulnerable), Fn. (Functional), Safe (Not Vulnerable)}}
\end{tabular}}
\vspace{-3mm}
\end{table}

\begin{table}[t]
\caption{The different hand-crafted vulnerable scenarios.\label{tab:scenarios-list}}
\resizebox{\columnwidth}{!}
{\begin{tabular}{@{}L{1.8cm}C{1cm}L{4cm}L{3cm}C{1cm}@{}}
\toprule
Scenario \newt{(`Top 25' Rank)} & Lang. & Description & Weakness & LoC \\ \midrule
CWE-79 \newt{(\#2)}  & Python   & A Flask blueprint for a web component to construct a simple welcome page.                             & Does not protect against cross-site scripting.                                          & 16  \\ \midrule
CWE-125 \newt{(\#3)} & C        & A function which should return a value from an array of integers or -1 if the index is out of bounds. & Does not check the lower bound.                                                         & 20  \\ \midrule
CWE-20  \newt{(\#4)} & Python   & A Flask blueprint to redirect users to a given destination URL on the `example.com' domain.           & Vulnerable to a prefix-domain name attack (e.g. redirect to `compromised-example.com'). & 14  \\ \midrule
CWE-416 \newt{(\#7)} & C        & Perform an XOR encryption of two strings.                                                             & A use-after-free bug.                                                                   & 49 \\ \midrule
CWE-476 \newt{(\#15)} & C        & A program to calculate the value of a person's account given assets and debts.                        & A null pointer exception when person name is unsuitable.                                & 42  \\ \midrule
CWE-119 \newt{(\#17)} & C        & A program which, given an ID number, returns a vehicle ID for that number from an array.              & Does not check any bounds.                                                              & 15  \\ \midrule
CWE-732 \newt{(\#22)} & C        & A program which writes a secret to a protected file.                                                  & Does not set appropriate permissions on the file when creating.                         & 23  \\ \bottomrule
\end{tabular}}
\end{table}

\changed{\textbf{Results}: \autoref{tbl:synthetic-vulnerable-repair} summarizes the synthetic program repair results. 2.2\% of the 22034 valid programs for the CWE-787 scenario were repaired. 29.6\% of the 10796 valid programs for the CWE-89 scenario were repaired.} 

\textbf{Our Observations:} From these two experiments (\autoref{tbl:synthetic-vulnerable-repair}) we can observe that the (Codex) \acp{LLM} \textit{are} capable of repairing vulnerable programs (RQ1) when given a suitable repair prompt. \changed{CWE-787 scenario saw 491 suggested patches that fix the bug (2.2\% of suggestions), and CWE-89 saw 3197 (29.6\%).}
\changed{\textit{In addition}, given that the original programs were also generated by the language model without bugs, it is also worth noting that the \acp{LLM} are capable of generating bug-free code even without additional context---assuming that they `understand' the generation task  (i.e., it is simple, like the original prompts in \autoref{fig:lst:synthetic-case-studies}).}
\changed{Indeed, the original generation task for the two scenarios gave fewer buggy programs (4.4\% of the programs were bug-free for CWE-787; 93.6\% for CWE-89) than the aggregated repair prompts (2.2\% and 29.6\% of patches were bug-free respectively). This indicates that while both scenarios were repaired by this prompt, the contents of the repair prompt should be further investigated (RQ2).}
\newt{\changed{Further, g}iven the results presented in \autoref{fig:iterative-generation-results}, where higher \textit{temperatures} are better for CWE-787 and worse for CWE-89 and vice versa, we can conclude that there is no single \textit{temperature}/\textit{top\_p} that will best cover all scenarios and engines (perhaps to be expected). However, given that the official Codex documentation advises varying either one of these parameters but not both, we can choose to set as parameters for an \textit{ensemble} of queries a set of temperatures $\{0.00, 0.25, 0.50, 0.75, 1.00\}$ with a \textit{top\_p} fixed to $1.00$. This prunes the ``search space'' for potential repairs by 80\% in our subsequent experiments.

}

\subsection{Prompt Engineering and Hand-Crafted Vulnerable Code\label{sec:handcraft-repair}}
For our next experiments, we dig deeper into RQ2, by (i) increasing the variety of prompts (increasing and decreasing the amount of context in the comments); and (ii) examining a wider \newt{and more complex} range of scenarios. Further, instead of generating scenarios synthetically via Codex (which may have follow-on implications upon the generated code) we hand-write short bespoke programs containing security weaknesses from MITRE's ``Top 25''~\cite{the_mitre_corporation_mitre_2021_2021} list.
As an additional point of comparison, we expand our analysis to include  other LLMs---the polycoder and gpt2-csrc models, which we ran locally; and AI21's `j1-large' and `j1-jumbo' (although unfortunately, due to their comparatively lower API usage limit, we were forced to reduce our sampling for these by 50\%).

\textbf{\newt{CWE choice and scenario design}:} Experimental scenarios are summarized in \autoref{tab:scenarios-list}.
\newt{They are designed from a selection of CWEs chosen for reasons similar to those in \autoref{sec:vuln-program-generation}: they are high-impact (in the `Top 25' list by MITRE~\cite{the_mitre_corporation_mitre_2021_2021}), concrete (straightforward to test presence/absence), and varied, with ``higher-level'' bugs inspired by Python web development and ``lower-level'' bugs being C buffer and pointer scenarios. These CWEs are a subset of those analyzed by automated tools in \cite{pearce_empirical_2021}, and our scenarios are inspired from their prior work.}

\textbf{\newt{Repair prompt design}:} After analysis by CodeQL, which finds our deliberately-inserted bugs, our experimental framework augments each scenario with a number of different possible \textit{prompt comment templates} to form the repair prompts (\autoref{fig:framework}).  As there are many possible wordings and language choices for constructing a repair prompt, we use guidance from user guides, informal searches of GitHub, and existing literature (see \autoref{sec:prompts}) to design five reasonable templates. 
These templates vary the amount of context provided to the LLM, \newt{from no information provided to extensive comments and hints}, including slight variations of the words and word order (e.g., `fixed' vs. `bugfix'), and including/excluding the faulty code. The templates are summarized in \autoref{tab:synthetic-templates}.

\begin{table}[]
\caption{Templates for repair prompt generation.}
\label{tab:synthetic-templates}
\resizebox{\linewidth}{!}{%
\begin{tabular}{@{}L{1.5cm}L{10cm}@{}}
\toprule
Template ID   & Description                                                                                                                                                                                                                                                                                          \\ \midrule
\texttt{n.h.} & No Help - deletes the vulnerable code/function body and provides no additional context for regeneration.                                                                                                                                                                                                           \\  \midrule
\texttt{s.1}  & Simple 1 - deletes the vulnerable code/function body and adds a comment `\texttt{bugfix: fixed [error name]}'.                                                                                                                                                                                          \\ \midrule
\texttt{s.2}  & Simple 2 - deletes the vulnerable code/function body and adds a comment `\texttt{fixed [error name] bug}'.                                                                                                                                                                                              \\ \midrule
\texttt{c.}   & Commented Code - After a comment `\texttt{BUG: [error name]}', this reproduces a `commented-out' version of the vulnerable code/function body followed by a `\texttt{FIXED:}'. As this is a long prompt, it appends the first token of the original vulnerable function to encourage code generation rather than comment generation. \\ \midrule

\texttt{c.m.} & Commented Code with Message - As with \texttt{c.}, but also includes a comment `\texttt{MESSAGE: [error message]}' and changes `\texttt{FIXED}' to `\texttt{FIXED VERSION}'. This style was used, without first token, in the earlier temperature sweep (see \autoref{fig:lst:string-case-study-bugfix-prompt}). \\ \midrule \midrule

\texttt{c.a.} & Commented Code (alternative) - \textit{Used for real-world examples, see \autoref{sec:realworld})}. As with \texttt{c.}, but commented in the alternative style for C programs (i.e., in C commenting, \texttt{/*} and \texttt{*/} rather than \texttt{//}).  \\ \midrule
\texttt{c.n.} & Commented Code (alternative), no token - \textit{Used for real-world examples, \autoref{sec:realworld})}. As with \texttt{c.a.}, but with no `first token' from vulnerable code. \\ \bottomrule
\end{tabular}}
\end{table}

\begin{figure}[]
\centering
    \includegraphics[width=0.9\linewidth]{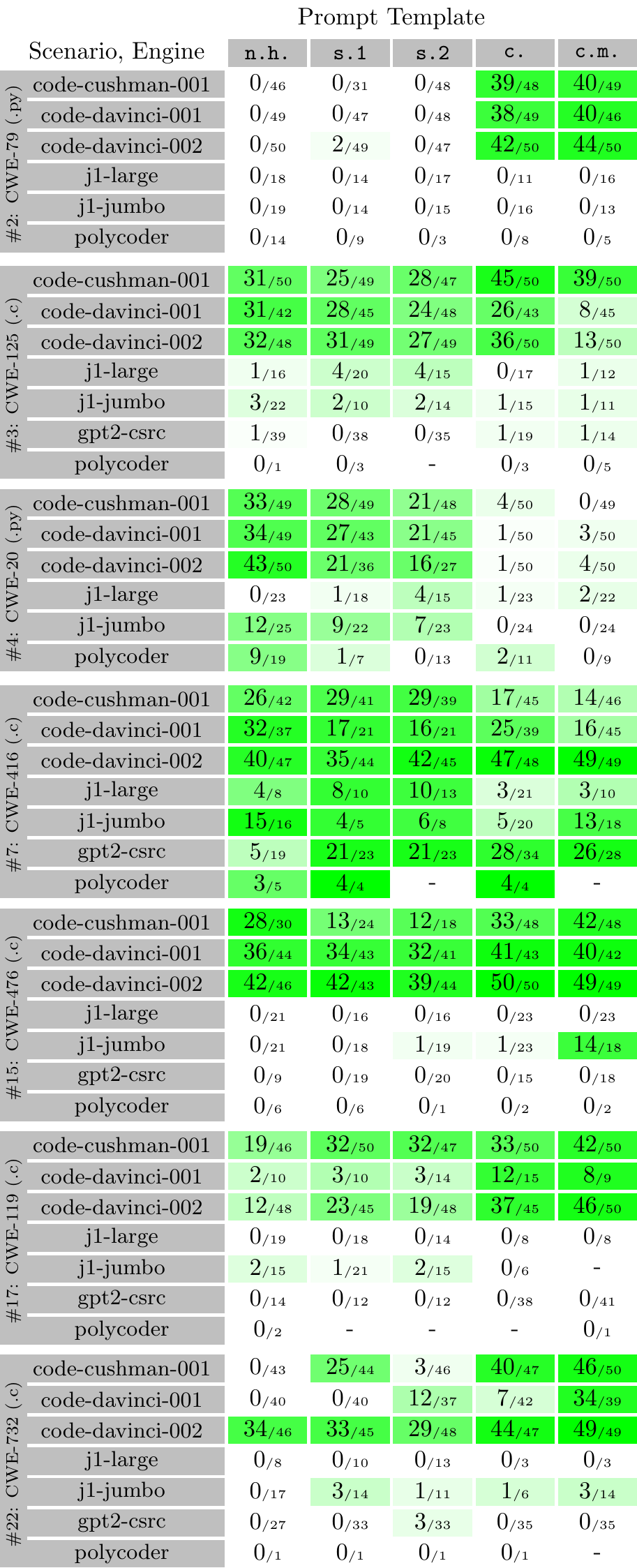}
    \caption{Results when using black-box \acp{LLM} to patch synthetically created bugs. Scenarios derived from the `Diversity of Weakness' dataset in \cite{pearce_empirical_2021}. Results are sorted by MITRE ``Top 25 CWE'' Rank \#. 10 programs (5 for AI21 models) were requested for each \textit{temperature} $\{0.00, 0.25, 0.50, 0.75, 1.00\}$ $\times$ \textit{top\_p} $\{1.00\}$, giving 50 (25 for AI21) possible programs. Results are presented as `safe and functional'/`valid (compiling) programs'.}
    \label{fig:synthetic-results-heatmaps}
\end{figure}

\begin{figure}
\centering
\begin{subfigure}[b]{\linewidth}
\begin{lstlisting}[language=python]
bp = Blueprint('hello',__name__)

#function which says hello to the user
@bp.route("hello/<username>")
def hello(username):
    return "<p>Hello, %s!</p>" % username
\end{lstlisting}
\vspace{-3mm}
\caption{Original vulnerable program (shortened for brevity)}
\label{fig:lst:xss-vulnerable-program-original}
\end{subfigure}

\vspace{1mm}
\begin{subfigure}[b]{\linewidth}
\begin{lstlisting}[language=python, linebackgroundcolor={\ifnum \value{lstnumber} = 4 \color{yellow} \fi}]
#function which says hello to the user
@bp.route("hello/<username>")
def hello(username):
    return "Hello, %s!" %escape(username)
\end{lstlisting}
\vspace{-3mm}
\caption{Template \texttt{n.h.} returns a safe program that fails functional tests, as the prompt (lines 1-3) does not indicate (`$<$p$>/<$p$>$' tags) are required.}
\label{fig:lst:xss-vulnerable-program-gen-bad}
\end{subfigure}

\vspace{1mm}
\begin{subfigure}[b]{\linewidth}
\begin{lstlisting}[language=python, linebackgroundcolor={\ifnum \value{lstnumber} = 7 \color{yellow} \fi}]
#function which says hello to the user
@bp.route("hello/<username>")
def hello(username):
    # BUG: Reflected server-side cross-site scripting
    #     return "<p>Hello, %s!</p>" % username
    # FIXED:
    return "<p>Hello, %s!</p>" % escape(username)

\end{lstlisting}
\vspace{-3mm}
\caption{Template \texttt{c.} returns a safe and functional program.}
\label{fig:lst:xss-vulnerable-program-gen-good}
\end{subfigure}

\vspace{-3mm}
\caption{CWE-79 (Python component with XSS vulnerability) program repair. Highlighted code was generated by the \acp{LLM}.}
\label{fig:lst:cwe-79-xss}
\vspace{-3mm}
\end{figure}

\begin{table}[t]
\caption{Hand-crafted program repair: template performance. Higher Valid Repair \% (i.e. `\# Fn. \& Safe' / `\# Vld.') are better.}
\label{tab:synthetic-template-performance}
\resizebox{\linewidth}{!}{%
\begin{tabular}{@{}llllllll@{}}
\toprule
\begin{tabular}[c]{@{}l@{}}Template \\ ID\end{tabular} & \# Gen. & \# Vld. & \# Vuln. & \# Fn. & \begin{tabular}[c]{@{}l@{}}\# Fn.\\ \& Vuln.\end{tabular} & \begin{tabular}[c]{@{}l@{}}\# Fn.\\ \& Safe\end{tabular} & \begin{tabular}[c]{@{}l@{}}\% Vld.\\ Repair\end{tabular} \\ \midrule
\texttt{n.h.}     & 2000    & 1316     & 340      & 646    & 116                                                       & 530                                                       & 40.2                                                      \\
\texttt{s.1}      & 2000    & 1213     & 247      & 539    & 94                                                       & 445                                                       & 36.7                                                     \\
\texttt{s.2}      & 2000    & 1204     & 315      & 592    & 126                                                       & 466                                                       & 38.7                                                     \\
\texttt{c.}       & 2000    & 1345     & 561      & 1140    & 475                                                       & 665                                                      & 49.4                                                    \\
\texttt{c.m.}     & 2000    & 1315     & 478      & 1104    & 414                                                       & 690                                                      & 52.5                                                     \\ \bottomrule
\multicolumn{8}{c}{{\footnotesize Gen. (Generated), Vld. (compilable), Vuln. (Vulnerable), Fn. (Functional), Safe (Not Vulnerable)}}
\end{tabular}
}
\vspace{-3mm}
\end{table}

\textbf{Results}: The results of this study are depicted in \autoref{fig:synthetic-results-heatmaps}, with the performance of each template presented collectively in \autoref{tab:synthetic-template-performance} and the total number of functional and safe programs generated by each engine presented in \autoref{fig:synthetic-results-heatmaps-totals-fig6} (in the Appendix).

\textbf{Our Observations:}
It is difficult to draw a definitive conclusion: the performance widely varied between prompt, scenario, and LLM. 
However, any one correct code completion is all that is required to fix a given bug, and \textbf{all scenarios were successfully repaired by at least one combination of template and engine}. 
The performance of each individual template can guide future work towards making a robust general-purpose single prompt (RQ2). 
\changed{In some cases, such as CWE-20, the low-context templates (\texttt{n.h}, \texttt{s.1}, and \texttt{s.2}) outperform the high-context templates (\texttt{c.} and \texttt{c.m.}), matching the results from the previous section (where original program generation performed better than program repair). However, in some scenarios, such as CWE-79 and CWE-732, this is reversed, and the high-context templates perform better.
When we perform a qualitative analysis of the results, we begin to understand this discrepancy. 
For many `low-context' templates, the generated prompt may not provide enough information to successfully generate code that will pass functional testing.
That is, the intent/functionality of the removed code is not adequately explained by what remains.
As such, even though the generated code may pass security tests, it fails to pass functional tests.
This is the case with CWE-79, for instance, which is reproduced in \autoref{fig:lst:xss-vulnerable-program-original}.
With a low-context template, the \ac{LLM} is often unable to determine the output format for the `hello' message without the context the commented out code provides (see \autoref{fig:lst:xss-vulnerable-program-gen-bad}). 
This is because once the faulty line is removed, there is no information provided to describe what the output \textit{should} look like as the comment describing the function is too vague.
However, when the additional context is provided (e.g. via high-context templates \texttt{c.} and \texttt{c.m.}), where the vulnerable code is instead `commented out' (see \autoref{fig:lst:xss-vulnerable-program-gen-good}), more information is made available to the \ac{LLM}, and now outputs may be generated correctly.}
\changed{This helps to explain the aggregated results in \autoref{tab:synthetic-template-performance}, where the high-context templates produced the best results on average.}
\changed{We believe that the additional technical detail provided helps \acp{LLM} to generate code to pass both security and functional tests for more difficult scenarios while still being useful for simpler bug-fixes. As such, we believe that this indicates that a more robust prompt should include more details rather than fewer.}

The OpenAI Codex models consistently outperform the other models with regards to generating successful patches. 
Given the apparent value in verbose prompts, we hypothesize that this relative performance is due to the broad training data in the original Codex models (i.e., based on GPT-3 models which were trained over English text, potentially leading to a better `understanding' of the comments). Nevertheless, all models, including our own `gpt2-csrc' were able to repair at least some of the programs. %

\subsection{Repairing Hardware CWEs\label{sec:diversityofdomain}}
As another component of our initial characterization of \acp{LLM}, we consider a more esoteric code completion setting: code written in \textit{hardware design languages} such as Verilog (which was explored in prior work~\cite{pearce_empirical_2021}). 
We now evaluate LLMs' ability to fix scenarios involving hardware CWEs. 
Buggy Verilog code may be used to undermine the confidentiality and/or integrity of the designed hardware.

\textbf{\newt{CWE choice and scenario design:}}
To examine the LLM performance on hardware, we designed two buggy hardware scenarios into our experimental data set. \newt{These scenarios are based on two Hardware CWEs which were chosen due to their relative straightforwardness: like the software CWEs examined above, their presence or absence within a snippet of Verilog code can be relatively easily determined. In addition, the code that is generated is relatively simple, which is important for these LLMs that have not been trained on extensive quantities of Verilog code. For similar reasons, these CWEs were also among those examined by the authors in \cite{pearce_empirical_2021}.} %
The first hardware CWE we examine is 
\textbf{CWE 1271}: Uninitialized Value on Reset for Registers Holding Security Settings. This arises when security-relevant assets such as locked registers are not set to appropriate values upon reset. 
The second is 
\textbf{CWE 1234}: Hardware Internal or Debug Modes Allow Override of Locks, which arises when a security-relevant asset in a locked register which should only be modifiable (when it is unlocked) is accessible or can be modified during \textit{debug} mode.

To evaluate these CWEs, we adapt code from examples on the MITRE website~\cite{the_mitre_corporation_mitre_cwe-1194_2021}. Our scenario for CWE-1271 is 14 lines of Verilog code and 17 lines for CWE-1234. We pass these source files to Verilator~\cite{noauthor_verilator_nodate} for functional and security testing before using these outputs within the framework (\autoref{fig:framework}) with the prompt templates from \autoref{tab:synthetic-templates}.

\begin{figure}
    \centering
    \includegraphics[width=0.9\linewidth]{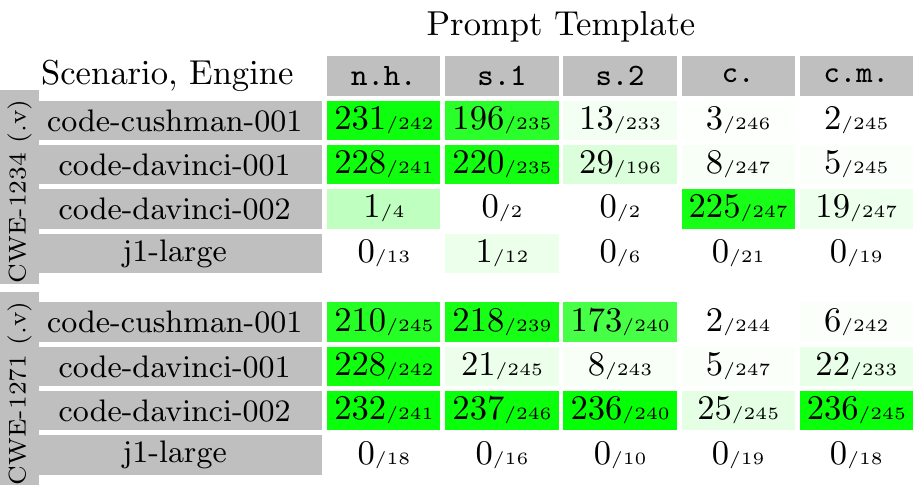}
    \caption{Results for \acp{LLM} when tasked with repairing two scenarios derived from the `Diversity of Domain' dataset in \cite{pearce_empirical_2021}. For each engine/prompt, (1) 10 of each combination of \textit{temperature} $\{0.00, 0.25, 0.50, 0.75, 1.00\}$ and \textit{top\_p} $\{0.00, 0.25, 0.50, 0.75, 1.00\}$, giving 250 possible total programs; (5 of each combination of \textit{temperature} $\{0.00, 0.25, 0.50, 0.75, 1.00\}$ and \textit{top\_p} of 1.00 for AI21's `j1-large', giving 25 possible programs), (2) the results are presented as `safe and functional'/`valid (synthesizing) programs'. }
    \label{fig:hw-heatmap}
    \vspace{-4mm}
\end{figure}

\textbf{Results:} Results are presented in \autoref{fig:hw-heatmap}. 

\textbf{Our Observations:}
Empirically, we found that LLMs were less proficient at producing Verilog code than they were at C or Python, so we include a complete sweep of \textit{temperatures} and \textit{top\_p} for the Codex models. 
Due to API usage restrictions we did not sweep the Jurrasic models. We also exclude `polycoder' and `gpt2-csrc' as they do not support Verilog.

\begin{figure}[t]
    \centering
    \includegraphics[width=0.8\linewidth]{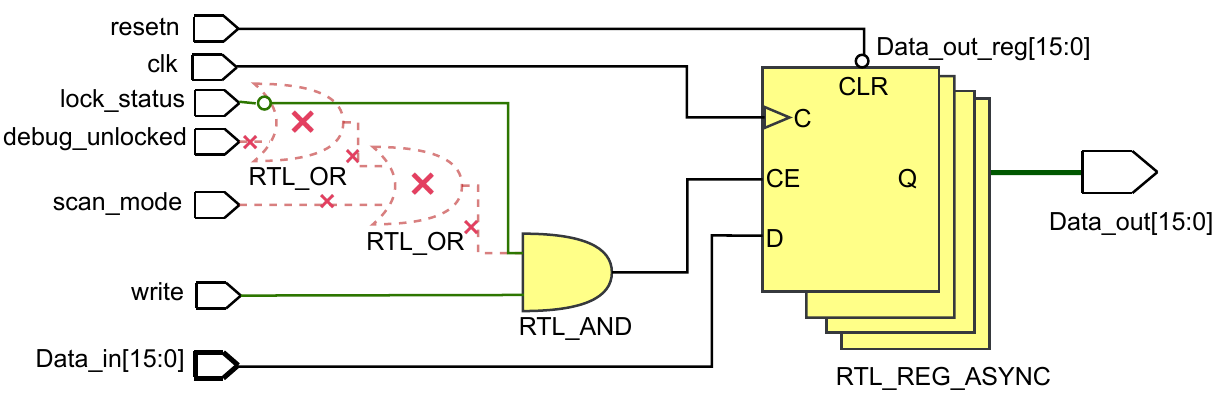}
    \caption{Schematic illustrating CWE-1234. Dotted red gates and wires are the added vulnerabilities. A `repair' involves removal of these elements such that the `clock enable' port of the \texttt{Data\_out} register is not dependent on the \texttt{debug\_unlocked} and \texttt{scan\_mode} signals.}
    \label{fig:cwe-1234-fig}
    \vspace{-4mm}
\end{figure}

To increase the \ac{LLM} code generation success, we add a post-processing step to Verilog generation. This is based around simple string analysis to add missing/remove redundant \texttt{end} and \texttt{endmodule} keywords.
For functional testing, we simulate the generated modules with a range of inputs using Verilator testbenches for each scenario. 
For CWE-1271, this involves testing a lock register is locked on reset. For CWE-1234, we test that a lock register properly controls access to a data register.
An example of correct and incorrect program synthesis is provided in~\autoref{fig:cwe-1234-fig}. Here, the incorrect design has a bug whereby the lock register can be overwritten when the \texttt{debug\_unlocked} or \texttt{scan\_mode} signals are high. The fix involves removing this logic, such that the register no longer has any dependency upon them.

The hardware repair examples returned somewhat different behavior as compared to the software repair. As seen in \autoref{fig:hw-heatmap}, the LLMs seemed perform better with less context provided in the prompt. In some sense, this can be thought of as the models having a tendency to `do the right thing.' 
As these two hardware components are conceptually simple with straightforward specifications, it may be that having any errors in the design files is simply less probable than having bug-free versions.

\section{RQ3: Experiments with Real-world CVEs}
\label{sec:realworld}

\subsection{Overview}

To further understand the capabilities (and limits) of \acp{LLM} for code repair, we now investigate real-world scenarios (CVEs) from public open-source projects (RQ3). 
This introduces several benefits and challenges.
Primarily, it allows us to characterize \ac{LLM} performance for much larger and much more realistic software projects.
The key challenge here is that we can no longer provide `full' context in a source file to the LLMs---whereas previously the entire vulnerable program could fit in the token limit of each LLM, real programs tend to be much than what these models can ingest. 
As such, we have to introduce a pre-generation \textit{reduction} step to fit the prompt into the token limit.

\subsection{ExtractFix Dataset\label{sec:extract-fix-explanation}}

\begin{table}[t]
\caption{Real World Scenarios (Subset of ExtractFix~\cite{gao_beyond_2021})}
\label{tab:real-world-scenario}
\resizebox{\linewidth}{!}{%
\begin{tabular}{@{}lp{3cm}llll@{}}
\toprule
Program & Description & File        & LoC   & EF\#     & CVE        \\ \midrule
\multirow{7}{*}{Libtiff}       & \multirow{7}{*}{\parbox{3cm}{\raggedright C library for processing TIFF files.}}     & tiffcrop.c & 9.1k  & EF01 & 2016-5321  \\
        &             & thumbnail.c & 683   & EF02-1 & 2014-8128  \\
        &             & tif\_next.c & 162   & EF02-2 & 2014-8128  \\
        &             & tiff2pdf.c  & 5.5k  & EF07   & 2016-10094 \\
        &             & tif\_jpeg.c & 2.3k  & EF08   & 2017-7601  \\
        &             & rgb2ycbcr.c & 393   & EF09   & 2016-3623  \\
        &             & tif\_jpeg.c & 2.4k  & EF10   & 2017-7595  \\ \midrule
\multirow{3}{*}{Libxml2}       & \multirow{3}{*}{\parbox{3cm}{XML C parser and toolkit.}}              & parser.c   & 15.7k & EF15 & 2016-1838  \\
        &             & parser.c    & 15.4k & EF17   & 2012-5134  \\
        &             & valid.c     & 7k    & EF18   & 2017-5969  \\ \midrule
\multirow{2}{*}{Libjpeg-turbo} & \multirow{2}{*}{\parbox{3cm}{C library for manipulating JPEG files.}} & wrbmp.c    & 557   & EF20 & 2018-19664 \\
        &             & jdmarker.c  & 1.3k  & EF22   & 2012-2806  \\ \bottomrule
\end{tabular}%
}
\vspace{-3mm}
\end{table}

We collected 12 real-world vulnerabilities across three projects drawn from the ExtractFix dataset used in prior work on program repair~\cite{gao_beyond_2021}. We chose the vulnerabilities based on (i) whether we could find a proof-of-concept input that triggered the vulnerability; (ii) whether the developer-provided patch was localized to a single file (since existing language models generally cannot consider multiple files' worth of context at once); and (iii) whether the project had a reasonably comprehensive test suite (for example, although ExtractFix contains two vulnerabilities in Jasper, a JPEG-2000 parser, Jasper did not have a test suite at the time these vulnerabilities were discovered). We note that restricting our evaluation to short, localized patches does not necessarily harm the validity of our analysis, as prior work has found that security patches tend to be more localized, have fewer source code modifications, and tend to affect fewer functions, compared to non-security bugs~\cite{li_large-scale_2017}.

To prepare each vulnerability for repair with our framework, we had to (i) Identify the patch (via its git commit hash) that fixed the vulnerability; (ii) Locate its parent commit, which identifies the version of the project that contains the vulnerability; (iii) Locate a proof-of-concept input that triggered the vulnerability; (iv) Build the project with the appropriate sanitizer (ASAN for memory safety issues and UBSAN for integer overflows, divide by zero, etc.); and (v) Determine how to run the project's regression tests. With this information in hand, we can check out the vulnerable version of the project, build it with a sanitizer, and then attempt repair, using the sanitizer output as an oracle to test whether the vulnerability has been fixed and the regression tests to ensure that the fix does not break other functionality.

\autoref{tab:real-world-scenario} provides an overview of the real-world projects and associated vulnerabilities that we tried to fix. The number of lines of code (LoC) varies even when the file to be fixed is the same, as each CVE corresponds to a different version (commit) of the project.

\textbf{Bug Localizing\label{sec:oracle}}: 
Given our focus on characterizing \acp{LLM} in the zero-shot bug-fix setting, we used the developer-provided patches as an oracle to \emph{localize} each vulnerability, prompting the \acp{LLM} to generate their fixes at the place in the code where the original developers patched each flaw. 
We note that although root cause identification and localization for security vulnerabilities is an active area of research~\cite{yagemann_automated_2021,yagemann_arcus_2021,zeller_simplifying_2002,choi_isolating_2002,sahoo_using_2013,kasikci_failure_2015}, this question is orthogonal to the repair-based research questions we investigate in this work.
Should \acp{LLM} prove useful for vulnerability repair, such work could be used as part of a complete, automated, end-to-end vulnerability repair system.

\begin{figure}
    \centering
    \includegraphics[width=0.65\linewidth]{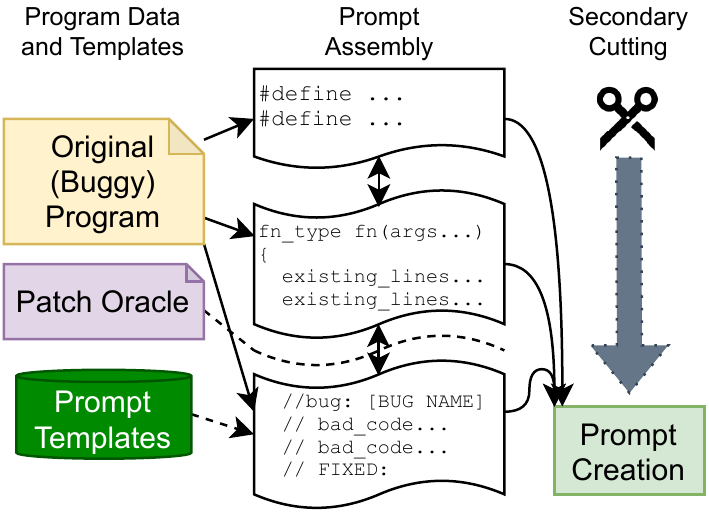}
    \caption{The prompt assembly extension for \autoref{fig:framework}}
    \label{fig:windowing-framework}
    \vspace{-6mm}
\end{figure}

\begin{figure}
\centering
\begin{subfigure}[b]{\linewidth}
\begin{lstlisting}[language=c,linebackgroundcolor={\ifnum\value{lstnumber}>8
            \ifnum\value{lstnumber}<10
                \color{pink}
            \fi
        \fi}]
/* Each tile contains only the data for a single plane
 * arranged in scanlines of tw * bytes_per_sample bytes.
 */
for (row = 0; row < imagelength; row += tl)
{ 
 nrow = (row + tl > imagelength) ? imagelength - row : tl;
 for (col = 0; col < imagewidth; col += tw)
 {
  for (s = 0; s < spp; s++)
  {  /* Read each plane of a tile set into srcbuffs[s] */
   tbytes = TIFFReadTile(in, srcbuffs[s], col, row, 0, s);
   if (tbytes < 0  && !ignore)
   {
    TIFFError(TIFFFileName(in),
    "Error, can't read tile for row %lu col %lu, "
...
\end{lstlisting}
\vspace{-3mm}
\caption{Buggy code from the large \texttt{tiffcrop.c} file. Error line 9 is shown in red.}
\label{fig:ef01-program-original}
\end{subfigure}

\begin{subfigure}[b]{\linewidth}
\begin{lstlisting}[language=c,linebackgroundcolor={\ifnum \value{lstnumber} = 7 \color{pink} \fi,\ifnum\value{lstnumber}>7
            \ifnum\value{lstnumber}<10
                \color{lightgray}
            \fi
        \fi}]
for (row = 0; row < imagelength; row += tl)
{ 
 nrow = (row + tl > imagelength) ? imagelength - row : tl;
 for (col = 0; col < imagewidth; col += tw)
 {
/* BUG: stack buffer overflow
 * for (s = 0; s < spp; s++)
 *  { // Read each plane of a tile set into srcbuffs[s] 
 * tbytes = TIFFReadTile(in, srcbuffs[s], col, row, 0, s);
 * FIXED:
 */
 for
\end{lstlisting}
\vspace{-3mm}
\caption{Prompt constructed according to \autoref{fig:windowing-framework}  (shortened for brevity). The red highlighted line 7 is the original faulty line indicated by ASAN/the oracle. The template includes lines 8 and 9 (highlighted in grey) to encourage the \acp{LLM} to regenerate the safe code so the patch can be matched safely.}
\label{fig:ef01-program-original1}
\end{subfigure}

\vspace{1mm}
\begin{subfigure}[b]{\linewidth}
\begin{lstlisting}[language=verilog,linebackgroundcolor={\ifnum \value{lstnumber} = 1 \color{yellow} \fi,\ifnum\value{lstnumber}>1
            \ifnum\value{lstnumber}<4
                \color{lightgray}
            \fi
        \fi}]
 (s = 0; (s < spp) && (s < MAX_SAMPLES); s++) 
 {
   tbytes = TIFFReadTile(in, srcbuffs[s], col, row, 0, s);
   /* END BUG FIX */
   if (tbytes < (tsize_t)(tw * nrow * bytes_per_sample)) 
   {
   TIFFError ("readSeparateTilesIntoBuffer", 
   ...
\end{lstlisting}
\vspace{-3mm}
\caption{Suggestion by code-cushman-001 which continues the prompt with the patch highlighted as yellow, line 1. Using the consolidation algorithm we match the gray highlighted lines 2 and 3 with the safe code line 11 and 12 from (b), allowing us to exclude the rest of the suggestion.} %
\label{fig:ef01-program-gen-gen}
\end{subfigure}

\vspace{1mm}
\begin{subfigure}[b]{\linewidth}
\begin{lstlisting}[language=c, linebackgroundcolor={\ifnum\value{lstnumber}>8
            \ifnum\value{lstnumber}<10
                \color{yellow}
            \fi
        \fi}]
/* Each tile contains only the data for a single plane
 * arranged in scanlines of tw * bytes_per_sample bytes.
 */
for (row = 0; row < imagelength; row += tl)
{ 
 nrow = (row + tl > imagelength) ? imagelength - row : tl;
 for (col = 0; col < imagewidth; col += tw)
 {
  for (s = 0; (s < spp) && (s < MAX_SAMPLES); s++)
  {
   tbytes = TIFFReadTile(in, srcbuffs[s], col, row, 0, s);
\end{lstlisting}
\vspace{-3mm}
\caption{The repaired program once reassembled with the \ac{LLM} patched line 11 highlighted in yellow. This generated patch is semantically equivalent with the real-world human patch used to repair this bug.}
\label{fig:ef01-program-gen-good}
\end{subfigure}

\vspace{-3.5mm}
\caption{Process to repair the real world bug in EF01 (see \autoref{tab:real-world-scenario}). This was the actual fix generated by code-davinci-001 (see \autoref{fig:lst:ef01-patch-dacinci-001} in the Appendix).}
\label{fig:list:full-repair-process-ef01}
\vspace{-6mm}
\end{figure}

\subsection{Code Reduction and Suggestion Consolidation}

The token limits for each model, noted in \autoref{tbl:LLMs}, functionally constrain the amount of code that (i) can be provided to the model, and (ii) can then be generated by the model (i.e., the number of tokens reflects both the input prompt and the output generated text). 
While these lengths were not an issue for the relatively short and simple programs in the hand-crafted examples from \autoref{tab:scenarios-list}, the real-world scenarios in \autoref{tab:real-world-scenario} are too large to be presented in their entirety.

\begin{figure*}[b]
\centering
\begin{subfigure}[b]{0.5\linewidth}
    \centering
    \includegraphics[width=0.95\linewidth]{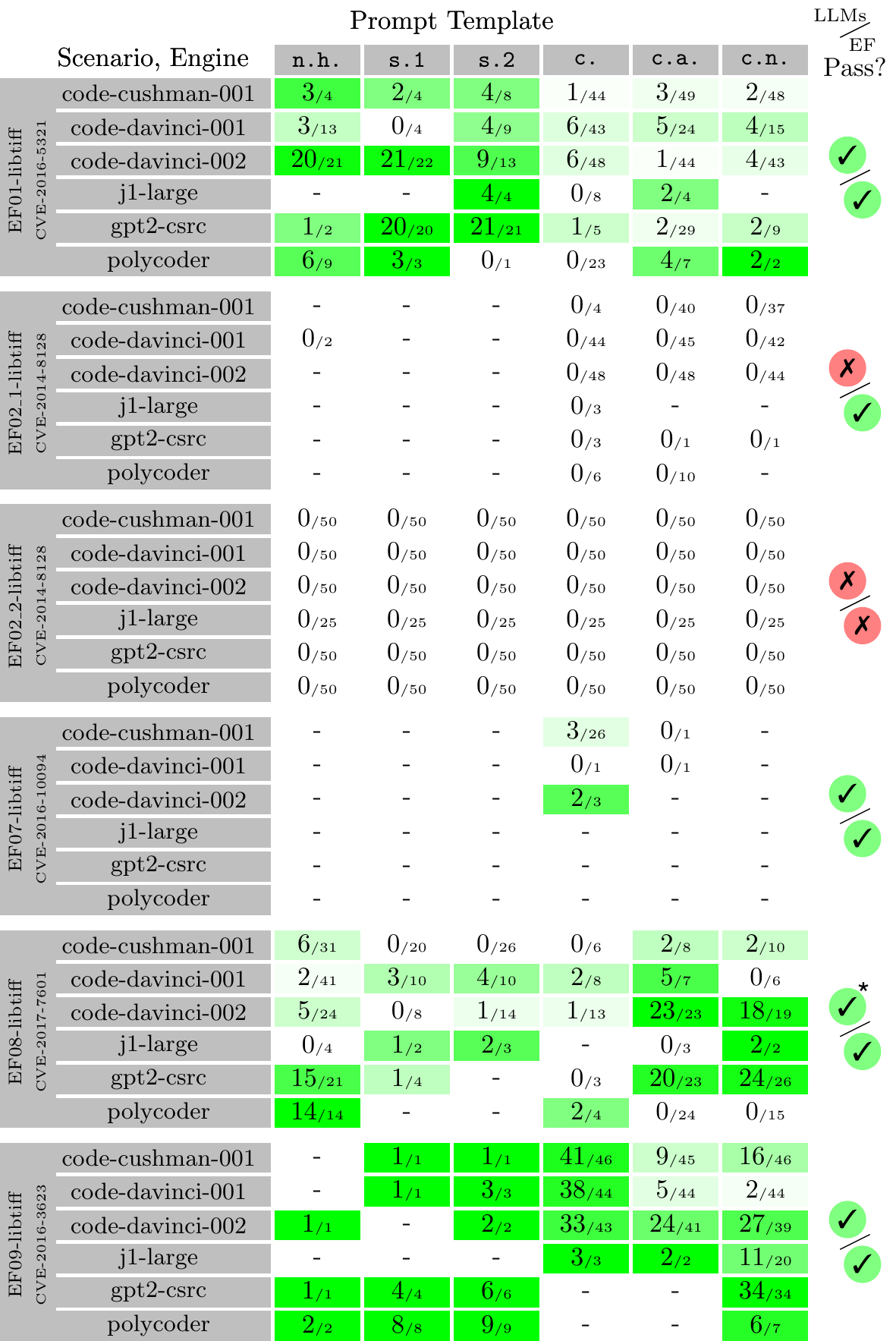}

\end{subfigure}%
\begin{subfigure}[b]{0.5\linewidth}
    \centering
    \includegraphics[width=0.95\linewidth]{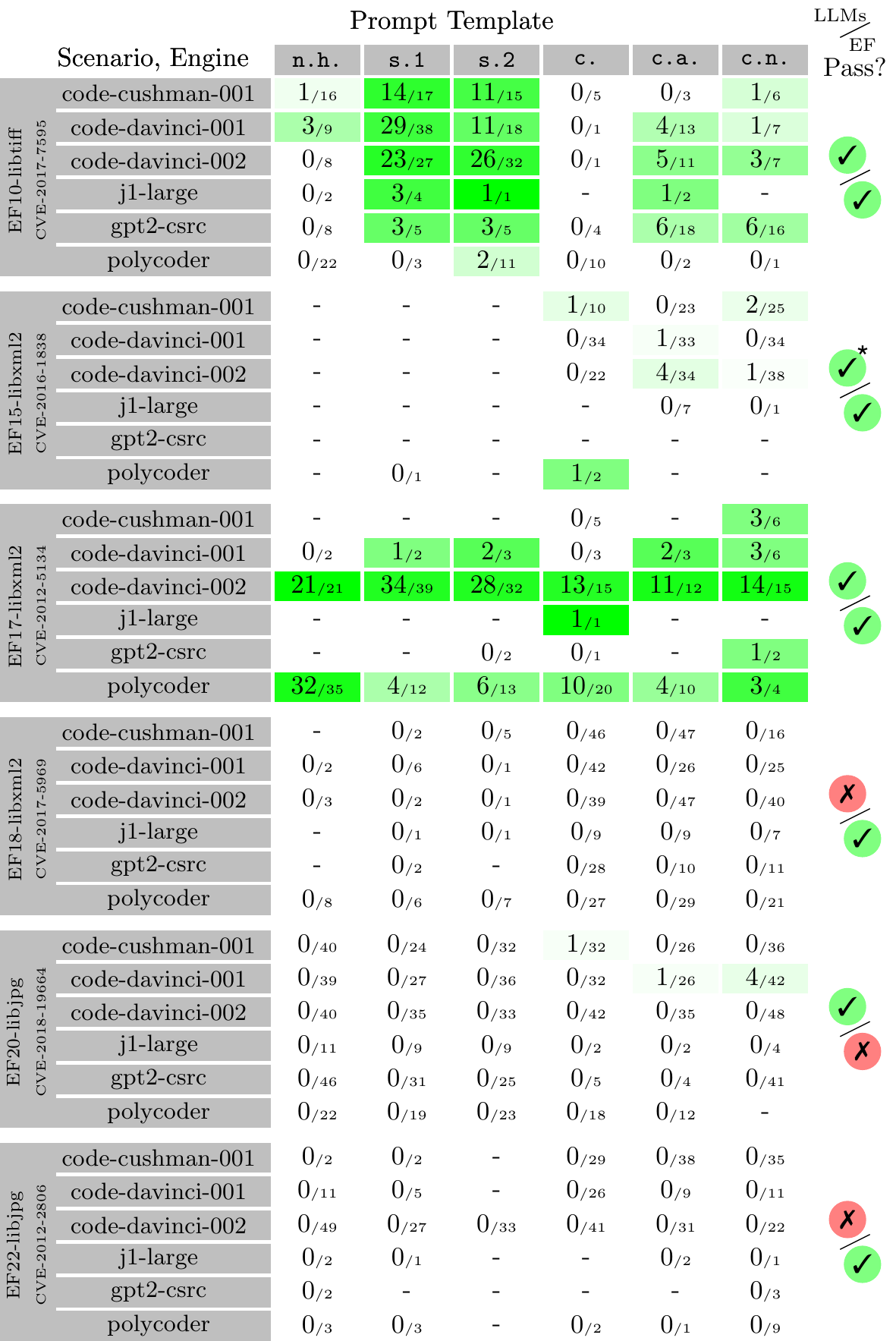}

\end{subfigure}
\caption{Results when using Black-box \acp{LLM} to patch real-world bugs from ExtractFix~\cite{gao_beyond_2021}. 10 programs (5 for AI21 model `j1-large') were requested for each \textit{temperature} $\{0.00, 0.25, 0.50, 0.75, 1.00\}$ $\times$ \textit{top\_p} $\{1.00\}$, giving 50 (25 for AI21) possible total programs. The results are presented as `safe and functional'/`valid (compiling) programs'. A scenario `passes' (is repaired) if any single response passes both functional and security testing. Results for `passing' presented for both LLMs (our work) and the original ExtractFix (EF) tool. \changed{\footnotesize *EF08 and *EF15 pass functional tests but are unreasonable patches. See \autoref{tab:res:plausible}.}}
\label{fig:realworld-results-heatmaps}
\vspace{-3mm}
\end{figure*}

\begin{figure}[t]
\vspace{-4mm}
\input{figs/diffs/ef07}
\vspace{-6mm}
\caption{\newt{`Successful' repair patches for EF07 with the highest confidence.}}
\label{fig:ef07-patches}
\end{figure}

To resolve this we extend the experimental framework in \autoref{fig:framework} to support code files larger than the token limit.
This extension consists of two parts, and is depicted in \autoref{fig:windowing-framework}. The first reduces the amount of code provided to the language model while still aiming to preserve enough context such that the bugs can be repaired. There is little existing advice on what this should include (see \autoref{sec:prompts}). 
Therefore, based on the user guides for the language models, which request context that `a developer would know', we begin the prompts by including the list of \texttt{\#define}s from the file. 
We then skip to the beginning of the function containing the vulnerable line(s), as defined by the \textit{oracle} (see \autoref{sec:extract-fix-explanation}). We then add the code as-is until the `buggy location', which is defined according to the oracle. We then include a comment template as in \autoref{sec:handcraft-repair}. 
This process alone may not suffice to fit the file into the LLM token limits. 
As such, we evaluate the length of the prompt by using each model's tokenizer (see \autoref{sec:codex}), and if it is too lengthy, we progressively cut lines from the top of the prompt until the prompt (and a statically-defined best-guess estimation of the number of tokens needed) fits into the token limit.

After the \ac{LLM} generates a response, we need to graft the new code with the existing file. 
We address this in three-stages. 
First, we attempt to find at least 30 characters of overlap in the LLM's response and the original file after vulnerable location (in the hope that the LLM produces a fix and then continues along the lines of the original code). 
To encourage this, we augment the prompt comment for the real-world scenario to include at least 2 lines of `safe' code along with the buggy lines (when using templates that include `commented-out' code).
Next, if no 30-character match is found, we progressively reduce the number of required characters and re-evaluate, to a minimum of 5 overlapping characters.
Finally, if no overlap is found at all, we insert the response from the \ac{LLM} up to the last generated new-line character (to prevent the addition of partial lines). 
We take the resultant code as LLM's fix suggestion and graft that into the original buggy file, between (i) where the original file was identified as buggy for the \ac{LLM} prompt generation, and (ii) where the oracle told us the original file was safe.

When manually inspecting their projects, we observed that single-line comments using \texttt{//} were uncommon. %
As such, we introduce a new template \texttt{c.a.} which adapts template \texttt{c.} to use the alternative notation (\texttt{/*} and \texttt{*/}) instead, and instead of using \texttt{c.m.} which relies on having a secondary `error message' which not be provided by the oracle we instead analyze a new variant \texttt{c.n.}.
This template is identical to \texttt{c.a.} but removes the `first token' of the vulnerable code.
These templates are listed in \autoref{tab:synthetic-templates}.
\autoref{fig:list:full-repair-process-ef01} presents a walk-through of the bug-patching process using the \texttt{c.a} template.

\subsection{Findings}

\textbf{Results:} The results for all prompts across all CVEs and \acp{LLM} are presented in \autoref{fig:realworld-results-heatmaps}. 
\changed{An LLM was able to repair 8 / 12 selected projects.  Note that we define a project `repaired' when the replaced code results in a compiled program passes both the functional tests included with each program and when it no longer crashes with the ASAN/UBSAN triggering input. }

\textbf{Our Observations:}
\changed{The ensemble's overall performance appears comparable to the state-of-the-art ExtractFix~\cite{gao_beyond_2021} repair tool which repaired 10 / 12.}
However, while many provided fixes do appear to fix the program (e.g., see the patches for EF07 in \autoref{fig:ef07-patches}), other `repaired' programs have patches that are \textit{implausible} (e.g. the patches for EF15, provided in \autoref{fig:ef15-patches} in the Appendix)\footnote{For interest we also include two other patches in the Appendix: EF01 (\autoref{fig:ef01-patches}) and EF20 (\autoref{fig:ef20-patches}). All other patches are in the open-source dataset.} -- while they `fix' the bugs and the programs pass their regression tests, they introduce other bugs which are not adequately tested for.
As it is not scalable to closely examine all `successful' programs, we manually examine the `highest-confidence' (as each LLM gives a relative `confidence score' with each output) patches that pass both functional and security testing in \autoref{tab:res:plausible}.
From this, we hypothesize that a majority of the `successful' programs may be unreasonable.
\changed{Noting that (i) the \acp{LLM} were not trained specifically for repair (i.e., this is a \textit{zero-shot} setting) and (ii) they can use only the limited context provided in their prompts, their performance is still remarkable. They even manage to convincingly fix one scenario (EF20) which ExtractFix could not.}

\begin{table}[t]
    \centering

    \caption{\newt{Author Opinions of LLM-provided Patches: \underline{Ident}ical or \underline{Sem}antically \underline{Eq}uivalent to the developer patch; \underline{R}easonable if they appear to fix the bug; or \underline{Not R}easonable if not.}}
    \label{tab:res:plausible}
\newt{
    \resizebox{0.9\linewidth}{!}{
    \begin{tabular}{lllllll}
\cline{1-3} \cline{5-7}
\textbf{Scenario}     & \textbf{Engine}  & \textbf{Plausibile} &  & \textbf{Scenario}     & \textbf{Engine}  & \textbf{Plausible} \\ \cline{1-3} \cline{5-7} 
\multirow{6}{*}{EF01} & code-cushman-001 & Not R.              &  & \multirow{6}{*}{EF10} & code-cushman-001 & R.                 \\
                      & code-davinci-001 & Sem. Eq.            &  &                       & code-davinci-001 & R.                 \\
                      & code-davinci-002 & Not R.              &  &                       & code-davinci-002 & R.                 \\
                      & j1-large         & Not R.              &  &                       & j1-large         & Not R.             \\
                      & gpt2-csrc        & Not R.              &  &                       & gpt2-csrc        & Not R.             \\
                      & polycoder        & Sem. Eq.            &  &                       & polycoder        & Not R.             \\ \cline{1-3} \cline{5-7} 
\multirow{2}{*}{EF07} & code-cushman-001 & Sem. Eq.             &  & \multirow{4}{*}{EF15} & code-cushman-001 & Not R.             \\
                      & code-davinci-002 & R.                  &  &                       & code-davinci-001 & Not R.             \\ \cline{1-3}
\multirow{6}{*}{EF08} & code-cushman-001 & Not R.              &  &                       & code-davinci-002 & Not R.             \\
                      & code-davinci-001 & Not R.              &  &                       & polycoder        & Not R.             \\ \cline{5-7} 
                      & code-davinci-002 & Not R.              &  & \multirow{6}{*}{EF17} & code-cushman-001 & Not R.             \\
                      & j1-large         & Not R.              &  &                       & code-davinci-001 & Ident.             \\
                      & gpt2-csrc        & Not R.              &  &                       & code-davinci-002 & Sem. Eq.           \\
                      & polycoder        & Not R.              &  &                       & j1-large         & Sem. Eq.            \\ \cline{1-3}
\multirow{6}{*}{EF09} & code-cushman-001 & R.                  &  &                       & gpt2-csrc        & Not R.             \\
                      & code-davinci-001 & R.                  &  &                       & polycoder        & Not R.             \\ \cline{5-7} 
                      & code-davinci-002 & R.                  &  & \multirow{2}{*}{EF20} & code-cushman-001 & R.                 \\
                      & j1-large         & Not R.              &  &                       & code-davinci-001 & Not R.             \\ \cline{5-7} 
                      & gpt2-csrc        & Not R.              &  &                       &                  &                    \\
                      & polycoder        & Not R.              &  &                       &                  &                    \\ \cline{1-3}
\end{tabular}
    }
    \vspace{-5mm}
}
\end{table}

As before, the performance across \acp{LLM} and prompts varies; the OpenAI models outperform the others.
Our `gpt2-csrc' is the `underdog';  it has far fewer parameters and is trained on a much smaller corpus (see \autoref{sec:gpt2-csrc}). Because we had access to the training data for this model, we checked whether any of the correct fixes it generated were included in the training set, and found that the fix for EF01 was indeed present. We hypothesize that the black-box \acp{LLM} might also benefit from this effect. %

The projects where \acp{LLM} failed have similar characteristics and tend to need large code changes (especially additions). Where new code is required (EF02\_1 and EF02\_2), the \acp{LLM} have limited `understanding' of the context to assist with the task. Meanwhile, EF18's real-world patch is long, removing 10 lines and adding 14: perhaps too onerous for \acp{LLM}. Finally, EF22 requires a  small patch. While it might seem that this could have been generated by the \acp{LLM}, the fix has tricky semantics; it alters the bounds of nested for loops, swapping arguments and adding a clause.

\newt{We present in \autoref{fig:realworld-results-heatmaps-totals} (in the Appendix) the sums of each template across each engine, and in \autoref{tab:realworld-template-performance} (in the Appendix) the per-template performance. Unlike in the synthetic examples, the \texttt{s.1} and \texttt{s.2} prompts were the most suitable. This could be because the UBSAN/ASAN messages are not as `helpful' as CodeQL messages, or because these longer programs might already have enough context. %
Because the bug classes are largely the same (C buffer errors) there is less diversity compared to the hand-crafted scenarios.}

We caution that many of the `successful' results noted in \autoref{fig:realworld-results-heatmaps} succeed only in the context of our definition of success: i.e., they pass the functional test suite for each project and no longer crash with an ASAN/UBSAN failure when given the original problematic input. %
Nevertheless, our experiments comprehensively characterize \acp{LLM} in a zero-shot setting.

\section{RQ4: Discussion on LLMs' Reliability}
\label{sec:dis}

\changed{Across 117 simple, synthetic programs for two CWEs, the \acp{LLM} generated 58,500 possible patches, of which 3,688 repaired their programs. We then hand-crafted 7 vulnerable programs to realize 7 CWEs. For these, the \acp{LLM} generated 10,000 possible patches, of which 2,796 repaired their programs (repairing 100\% of the programs). We used 12 real-world projects with historical CVEs and had the \acp{LLM} generate 19,600 possible patches, of which 982 patches `repaired' their programs (8-out-of-12).
Generally, detailed prompts are more effective in coaxing models towards producing patched programs. 
\newt{LLMs worked best when they only had to produce short, local fixes. Where complex context was required, they performed worse.} %

Taking into consideration the complete suite of experiments in our study, on the question "How reliable are LLMs at generating repairs?", we find the answer to be mixed. 
In general, we are surprised by the quality and success of the code generation by the \acp{LLM} when tasked with security bug repair,  considering that 100\% of the synthetic and hand-crafted scenarios were convincingly repaired.
However, based on the qualitative analysis of the real-world scenarios, we do not yet think the LLMs are sufficiently reliable to usurp automatic program repair. 
Further, other LLM-based constraints apply: repairs are restricted to a single place within a single file, and although security bugs tend to be more localized than other bugs~\cite{li_large-scale_2017}, this is not universal.}%

\section{Study Limitations \label{sec:limitations}}

\textbf{Potentially Inadequate Functional Tests.} 
A well-known problem in program repair is that regression tests for a project are  \emph{weak proxies} for the correctness of the program. 
For example, Qi et al.~\cite{qi_analysis_2015} found that although GenProg~\cite{le_goues_genprog_2012} claimed to fix 55 bugs, with stricter testing only two of the patches were correct. \newt{As noted in \autoref{tab:res:plausible}, this limitation applies to patches generated by evaluated \acp{LLM}}. 
If testing suites were not comprehensive, programs that appear `repaired' may not truly be so.
Future work should consider robust approaches to evaluation (e.g., fuzzing). 

\newt{
\textbf{Security Tool Test Dependency.} 
Evaluating the LLM patches also required pairing them with bug detection tools. For the initial experiments (synthetic and hand-crafted scenarios) we used CodeQL, which has also been used by others~\cite{pearce_empirical_2021}. To evaluate its performance, we performed extensive manual validation during the set-up phases of this project, and every bug that was repaired was built from bug reports generated by CodeQL itself (i.e., in \autoref{fig:framework} both `Security Tests' are implemented by CodeQL running the exact same queries). %
To minimize author-introduced bugs in the tooling, we relied on CodeQL's repository of existing queries. Of our software-based synthetic and hand-crafted scenarios, all could be analyzed by CodeQL.

For the real-world scenarios we rely on crashing inputs (in ASAN/UBSAN). We ensured that the input that caused the failing program behavior as reported by ASAN was tested (a) against the original program (to observe failing behavior), and (b) against the "repaired" program (to see if it still fails). A pass guarantees that the failure case is repaired, however does not guarantee that the vulnerability is repaired, as %
it is impossible for testing to prove a bug's absence.

}

\textbf{Scenario Design.} While we tried to capture a wide variety of different security weaknesses in different scenarios, there remain many languages and weaknesses not considered.

\textbf{Prompt Engineering.} While we designed several prompt templates, other approaches can be explored. We did not explore cases whereby multiple files needed to be patched (or comprehended) simultaneously. Deciding which context to provide given limited \acp{LLM} tokens is an open challenge.

\textbf{Vulnerability Disclosure.} We use synthetic, hand-crafted, and historic vulnerabilities, so no disclosure was required.

\section{Related Prior Work\label{sec:related}}
Studies in the design and maintenance of software systems has produced a vast body of research, such as insights into the nature of security bugs/weaknesses~\cite{the_mitre_corporation_mitre_2021_2021} and security patches. For example,
 security patches are more localized and with fewer source code modifications compared to non-security bugs~\cite{li_large-scale_2017}.
Other work has examined practices around comments in code~\cite{tan_icomment_2007,misra_is_2020,pascarella_classifying_2017}. 
Comments serve several purposes, from notices through to explanations of parts of the code for other developers~\cite{misra_is_2020,pascarella_classifying_2017}. 
While comments can provide useful context, they can also mislead developers~\cite{tan_icomment_2007}. 
Researchers studied commented-out code~\cite{pham_secret_2020}, where code is preserved (sometimes temporarily) over the course of debugging, adding functionality, and other stages of software design. 

Given increasing complexity of software and the ongoing pursuit of development process efficiency, researchers have investigated myriad ways to support software development, such as tools for code completion~\cite{austin_program_2021,chen_evaluating_2021,mihalcea_nlp_2006,drechsler_generating_2012,ai21_discover_nodate} or recommending fixes based on compiler error messages by ``crowdsourcing'' collective wisdom/experience~\cite{hartmann_what_2010,jeffrey_bugfix_2009}. 
Other prior work includes techniques and tools for detecting security bugs~\cite{owasp_source_nodate,github_inc_codeql_2021,bandara_fix_2020,li_fuzzing_2018}. 
Exploiting our collective experiences for helping software designers is, in a sense, exemplified by emerging \ac{ML}-based approaches including the LLMs we investigated~\cite{chen_evaluating_2021,ai21_jurassic-1_2021,lieber_jurassic-1_2021}, given their training on large quantities of open-source code. 
Such models have an enormous number of parameters and are offered as a black-box tool for software developers to try.  
To date, we are aware of only the study by Pearce et al.~\cite{pearce_empirical_2021} that applies a security lens to gauge the security implications of AI-generated code completions in a large, black-box product (GitHub Copilot), although without considering functional correctness. 

While this study investigates using \acp{LLM} to recommend fixes for security bugs, there is broader research in automatic software repair. For a survey, consider work by Monperrus~\cite{monperrus_automatic_2018}. 
Literature in this area deals with the repair of programs that violate a specification (e.g., as captured by a functional test suite), of which repairing \textit{security} bugs is one specialization~\cite{gao_beyond_2021}. 
Approaches include those based on symbolic execution and formal properties~\cite{gao_beyond_2021}, matching code to a database of bug-fix scenarios~\cite{jeffrey_bugfix_2009}, and \ac{NMT}-based approaches for bug fixes (e.g., \cite{tufano_empirical_2018, jiang_cure_2021,chen_sequencer_2021,lutellier_coconut_2020}). 

\ac{NMT}-based approaches train dedicated models, such as recurrent neural network (RNN)-based encoder-decoder architectures~\cite{tufano_empirical_2019} on explicit ``bug-fix pairs'' mined from software repositories (e.g., through finding code changes through version control logs). 
Tufano et al.'s approach learns patches for individual functions, with ``less than 100 tokens'' and demonstrated some potential in generating fixes (they reported 9.22\% on their test set). 
Their approach does not consider comments. 
DeepDebug~\cite{drain_generating_2021} requires the curation of bug/patch datasets and formulates the goal of predicting human-designed patches. 
An approach by Jiang et al. first pre-trains a model on a large software code-base to ``learn developer-like source code'' before fine-tuning to specialize on the program repair task~\cite{jiang_cure_2021}. 
They propose techniques to improve the search for viable patches amongst several predicted tokens. 
SequenceR~\cite{chen_sequencer_2021} focuses on ``one line fixes'' by curating a dataset of commits that modify single lines of code as well as a copy mechanism for working around the model architecture's limited vocabulary. 
\ac{NMT}-based approaches steer clear of problems due to ``overfitting'' test suites~\cite{qi_analysis_2015}, a pitfall of \textit{generate-and-validate} approaches for patch generation (e.g., the seminal GenProg~\cite{le_goues_genprog_2012}).
By setting the target for repair models to be  (re)-production of human-generated patches, the expectation is that human-generated patches are more likely to be \textit{correct} (leading to correct outputs) rather than \textit{plausible} (``passes'' test suites without guarantees of correct output). Human-generated security patches are not always correct~\cite{li_large-scale_2017}. 

Generally, prior work has restrictions (e.g., single line fixes or language-specificity). 
Hence, we investigated larger, ``general-purpose'' LLMs and tried to characterize their ability in the security bug fixing domain, without specialization. 
Our approach of probing the models as bug-fix recommenders, without fine-tuning, bears some similarity to studies of \acp{LLM} in different \ac{NLP} tasks~\cite{radford_language_2019}.

\section{Conclusions and Future Work\label{sec:conc}\label{sec:conclusions}}
In this paper we set out to characterize \acfp{LLM} for code in terms of their ability to repair software vulnerabilities in a zero-shot setting. 
\changed{We found that general purpose, black-box LLMs can generate fixes for security bugs when given a carefully constructed prompt, including in 100\% of our synthetic and hand-crafted scenarios.}
\changed{However, our evaluation of the LLM's performance shows that the state of the art is not yet enough for the approach to deliver real value in a program repair framework.}

\changed{While we believe future, targeted \acp{LLM} have potential for use in this area, challenges remain. For a full end-to-end system, complete systems will need to incorporate bug localization and improved test rigs for evaluation. 
Querying an ensemble of \acp{LLM} is unlikely to scale for large numbers of developers to adopt in their day-to-day workloads.}

\section*{Acknowledgments}
We would like to thank the anonymous reviewers for their constructive feedback and encouragement. 
This work was partially supported by ONR Award \# N00014-18-1-2058 and NSF award \# 2145482. 
Any opinions, findings, and conclusions, or recommendations expressed are those of the authors and do not necessarily reflect the views of the sponsors.

\clearpage
\bibliographystyle{IEEEtran}
\bibliography{lit/benhamram}

\begin{thebibliography}{10}
\providecommand{\url}[1]{#1}
\csname url@samestyle\endcsname
\providecommand{\newblock}{\relax}
\providecommand{\bibinfo}[2]{#2}
\providecommand{\BIBentrySTDinterwordspacing}{\spaceskip=0pt\relax}
\providecommand{\BIBentryALTinterwordstretchfactor}{4}
\providecommand{\BIBentryALTinterwordspacing}{\spaceskip=\fontdimen2\font plus
\BIBentryALTinterwordstretchfactor\fontdimen3\font minus
  \fontdimen4\font\relax}
\providecommand{\BIBforeignlanguage}[2]{{%
\expandafter\ifx\csname l@#1\endcsname\relax
\typeout{** WARNING: IEEEtran.bst: No hyphenation pattern has been}%
\typeout{** loaded for the language `#1'. Using the pattern for}%
\typeout{** the default language instead.}%
\else
\language=\csname l@#1\endcsname
\fi
#2}}
\providecommand{\BIBdecl}{\relax}
\BIBdecl

\bibitem{chen_evaluating_2021}
\BIBentryALTinterwordspacing
M.~Chen, J.~Tworek, H.~Jun, Q.~Yuan, H.~P. d.~O. Pinto, J.~Kaplan, H.~Edwards,
  Y.~Burda, N.~Joseph, G.~Brockman, A.~Ray, R.~Puri, G.~Krueger, M.~Petrov,
  H.~Khlaaf, G.~Sastry, P.~Mishkin, B.~Chan, S.~Gray, N.~Ryder, M.~Pavlov,
  A.~Power, L.~Kaiser, M.~Bavarian, C.~Winter, P.~Tillet, F.~P. Such,
  D.~Cummings, M.~Plappert, F.~Chantzis, E.~Barnes, A.~Herbert-Voss, W.~H.
  Guss, A.~Nichol, A.~Paino, N.~Tezak, J.~Tang, I.~Babuschkin, S.~Balaji,
  S.~Jain, W.~Saunders, C.~Hesse, A.~N. Carr, J.~Leike, J.~Achiam, V.~Misra,
  E.~Morikawa, A.~Radford, M.~Knight, M.~Brundage, M.~Murati, K.~Mayer,
  P.~Welinder, B.~McGrew, D.~Amodei, S.~McCandlish, I.~Sutskever, and
  W.~Zaremba, ``Evaluating {Large} {Language} {Models} {Trained} on {Code},''
  \emph{arXiv:2107.03374 [cs]}, Jul. 2021, arXiv: 2107.03374. [Online].
  Available: \url{http://arxiv.org/abs/2107.03374}
\BIBentrySTDinterwordspacing

\bibitem{github_github_nodate}
\BIBentryALTinterwordspacing
GitHub, ``{GitHub} {Copilot} {\textperiodcentered} {Your} {AI} pair
  programmer.'' [Online]. Available: \url{https://copilot.github.com/}
\BIBentrySTDinterwordspacing

\bibitem{ai21_discover_nodate}
\BIBentryALTinterwordspacing
AI21, ``Discover {Use} {Cases} for {AI21} {Studio} and {Jurassic}-1.''
  [Online]. Available: \url{https://www.ai21.com/blog/ai21-studio-use-cases}
\BIBentrySTDinterwordspacing

\bibitem{openai_completion_nodate}
\BIBentryALTinterwordspacing
OpenAI, ``\BIBforeignlanguage{en}{Completion - {OpenAI} {API}}.'' [Online].
  Available: \url{https://beta.openai.com/docs/guides/completion/prompt-design}
\BIBentrySTDinterwordspacing

\bibitem{brown_language_2020}
\BIBentryALTinterwordspacing
T.~B. Brown, B.~Mann, N.~Ryder, M.~Subbiah, J.~Kaplan, P.~Dhariwal,
  A.~Neelakantan, P.~Shyam, G.~Sastry, A.~Askell, S.~Agarwal, A.~Herbert-Voss,
  G.~Krueger, T.~Henighan, R.~Child, A.~Ramesh, D.~M. Ziegler, J.~Wu,
  C.~Winter, C.~Hesse, M.~Chen, E.~Sigler, M.~Litwin, S.~Gray, B.~Chess,
  J.~Clark, C.~Berner, S.~McCandlish, A.~Radford, I.~Sutskever, and D.~Amodei,
  ``Language {Models} are {Few}-{Shot} {Learners},'' \emph{arXiv:2005.14165
  [cs]}, Jul. 2020, arXiv: 2005.14165. [Online]. Available:
  \url{http://arxiv.org/abs/2005.14165}
\BIBentrySTDinterwordspacing

\bibitem{radford_language_2019}
\BIBentryALTinterwordspacing
A.~Radford, J.~Wu, R.~Child, D.~Luan, D.~Amodei, and I.~Sutskever,
  ``\BIBforeignlanguage{en}{Language {Models} are {Unsupervised} {Multitask}
  {Learners}},'' p.~24, 2019. [Online]. Available:
  \url{https://cdn.openai.com/better-language-models/language_models_are_unsupervised_multitask_learners.pdf}
\BIBentrySTDinterwordspacing

\bibitem{openai_openai_2021}
\BIBentryALTinterwordspacing
OpenAI, ``\BIBforeignlanguage{en}{{OpenAI} {Codex}},'' Aug. 2021. [Online].
  Available: \url{https://openai.com/blog/openai-codex/}
\BIBentrySTDinterwordspacing

\bibitem{lieber_jurassic-1_2021}
\BIBentryALTinterwordspacing
O.~Lieber, O.~Sharir, B.~Lentz, and Y.~Shoham,
  ``\BIBforeignlanguage{en}{Jurassic-1: {Technical} {Details} and
  {Evaluation}},'' AI21 Labs, Tech. Rep., Aug. 2021. [Online]. Available:
  \url{https://uploads-ssl.webflow.com/60fd4503684b466578c0d307/61138924626a6981ee09caf6_jurassic_tech_paper.pdf}
\BIBentrySTDinterwordspacing

\bibitem{misra_is_2020}
\BIBentryALTinterwordspacing
V.~Misra, J.~S.~K. Reddy, and S.~Chimalakonda, ``Is there a correlation between
  code comments and issues? an exploratory study,'' in \emph{Proceedings of the
  35th {Annual} {ACM} {Symposium} on {Applied} {Computing}}, ser. {SAC}
  '20.\hskip 1em plus 0.5em minus 0.4em\relax New York, NY, USA: Association
  for Computing Machinery, Mar. 2020, pp. 110--117. [Online]. Available:
  \url{https://doi.org/10.1145/3341105.3374009}
\BIBentrySTDinterwordspacing

\bibitem{pascarella_classifying_2017}
\BIBentryALTinterwordspacing
L.~Pascarella and A.~Bacchelli, ``Classifying {Code} {Comments} in {Java}
  {Open}-{Source} {Software} {Systems},'' in \emph{2017 {IEEE}/{ACM} 14th
  {International} {Conference} on {Mining} {Software} {Repositories}
  ({MSR})}.\hskip 1em plus 0.5em minus 0.4em\relax Buenos Aires, Argentina:
  IEEE, May 2017, pp. 227--237. [Online]. Available:
  \url{http://ieeexplore.ieee.org/document/7962372/}
\BIBentrySTDinterwordspacing

\bibitem{tan_icomment_2007}
\BIBentryALTinterwordspacing
L.~Tan, D.~Yuan, G.~Krishna, and Y.~Zhou, ``\BIBforeignlanguage{en}{/*icomment:
  bugs or bad comments?*/},'' \emph{\BIBforeignlanguage{en}{ACM SIGOPS
  Operating Systems Review}}, vol.~41, no.~6, pp. 145--158, Oct. 2007.
  [Online]. Available: \url{https://dl.acm.org/doi/10.1145/1323293.1294276}
\BIBentrySTDinterwordspacing

\bibitem{pearce_empirical_2021}
\BIBentryALTinterwordspacing
H.~Pearce, B.~Ahmad, B.~Tan, B.~Dolan-Gavitt, and R.~Karri, ``An {Empirical}
  {Cybersecurity} {Evaluation} of {GitHub} {Copilot}'s {Code}
  {Contributions},'' \emph{arXiv:2108.09293 [cs]}, Aug. 2021, arXiv:
  2108.09293. [Online]. Available: \url{http://arxiv.org/abs/2108.09293}
\BIBentrySTDinterwordspacing

\bibitem{chen_sequencer_2021}
Z.~Chen, S.~Kommrusch, M.~Tufano, L.-N. Pouchet, D.~Poshyvanyk, and
  M.~Monperrus, ``{SequenceR}: {Sequence}-to-{Sequence} {Learning} for
  {End}-to-{End} {Program} {Repair},'' \emph{IEEE Transactions on Software
  Engineering}, vol.~47, no.~9, pp. 1943--1959, Sep. 2021.

\bibitem{jiang_cure_2021}
N.~Jiang, T.~Lutellier, and L.~Tan, ``{CURE}: {Code}-{Aware} {Neural} {Machine}
  {Translation} for {Automatic} {Program} {Repair},'' in \emph{2021
  {IEEE}/{ACM} 43rd {International} {Conference} on {Software} {Engineering}
  ({ICSE})}, May 2021, pp. 1161--1173, iSSN: 1558-1225.

\bibitem{tufano_empirical_2019}
\BIBentryALTinterwordspacing
M.~Tufano, C.~Watson, G.~Bavota, M.~D. Penta, M.~White, and D.~Poshyvanyk, ``An
  {Empirical} {Study} on {Learning} {Bug}-{Fixing} {Patches} in the {Wild} via
  {Neural} {Machine} {Translation},'' \emph{ACM Transactions on Software
  Engineering and Methodology}, vol.~28, no.~4, pp. 19:1--19:29, Sep. 2019.
  [Online]. Available: \url{https://doi.org/10.1145/3340544}
\BIBentrySTDinterwordspacing

\bibitem{drain_generating_2021}
\BIBentryALTinterwordspacing
D.~Drain, C.~Wu, A.~Svyatkovskiy, and N.~Sundaresan,
  ``\BIBforeignlanguage{en}{Generating bug-fixes using pretrained
  transformers},'' in \emph{\BIBforeignlanguage{en}{Proceedings of the 5th
  {ACM} {SIGPLAN} {International} {Symposium} on {Machine}
  {Programming}}}.\hskip 1em plus 0.5em minus 0.4em\relax Virtual Canada: ACM,
  Jun. 2021, pp. 1--8. [Online]. Available:
  \url{https://dl.acm.org/doi/10.1145/3460945.3464951}
\BIBentrySTDinterwordspacing

\bibitem{li_large-scale_2017}
\BIBentryALTinterwordspacing
F.~Li and V.~Paxson, ``\BIBforeignlanguage{en}{A {Large}-{Scale} {Empirical}
  {Study} of {Security} {Patches}},'' in
  \emph{\BIBforeignlanguage{en}{Proceedings of the 2017 {ACM} {SIGSAC}
  {Conference} on {Computer} and {Communications} {Security}}}.\hskip 1em plus
  0.5em minus 0.4em\relax Dallas Texas USA: ACM, Oct. 2017, pp. 2201--2215.
  [Online]. Available: \url{https://dl.acm.org/doi/10.1145/3133956.3134072}
\BIBentrySTDinterwordspacing

\bibitem{the_mitre_corporation_mitre_cwe_2020}
\BIBentryALTinterwordspacing
T.~M.~C. (MITRE), ``{CWE} - {CWE}-{Compatible} {Products} and {Services},''
  Dec. 2020. [Online]. Available:
  \url{https://cwe.mitre.org/compatible/compatible.html}
\BIBentrySTDinterwordspacing

\bibitem{owasp_source_nodate}
\BIBentryALTinterwordspacing
OWASP, ``\BIBforeignlanguage{en}{Source {Code} {Analysis} {Tools}}.'' [Online].
  Available: \url{https://owasp.org/www-community/Source_Code_Analysis_Tools}
\BIBentrySTDinterwordspacing

\bibitem{serebryany_addresssanitizer_2012}
\BIBentryALTinterwordspacing
K.~Serebryany, D.~Bruening, A.~Potapenko, and D.~Vyukov, ``{AddressSanitizer}:
  {A} {Fast} {Address} {Sanity} {Checker},'' in \emph{2012 {USENIX} {Annual}
  {Technical} {Conference} ({USENIX} {ATC} 12)}.\hskip 1em plus 0.5em minus
  0.4em\relax Boston, MA: USENIX Association, Jun. 2012, pp. 309--318.
  [Online]. Available:
  \url{https://www.usenix.org/conference/atc12/technical-sessions/presentation/serebryany}
\BIBentrySTDinterwordspacing

\bibitem{polacek_gcc_2014}
\BIBentryALTinterwordspacing
M.~Polacek, ``\BIBforeignlanguage{en}{{GCC} {Undefined} {Behavior} {Sanitizer}
  - ubsan},'' Oct. 2014. [Online]. Available:
  \url{https://developers.redhat.com/blog/2014/10/16/gcc-undefined-behavior-sanitizer-ubsan}
\BIBentrySTDinterwordspacing

\bibitem{monperrus_automatic_2018}
\BIBentryALTinterwordspacing
M.~Monperrus, ``Automatic {Software} {Repair}: {A} {Bibliography},'' \emph{ACM
  Computing Surveys}, vol.~51, no.~1, pp. 17:1--17:24, Jan. 2018. [Online].
  Available: \url{https://doi.org/10.1145/3105906}
\BIBentrySTDinterwordspacing

\bibitem{gage_new_1994}
P.~Gage, ``A {New} {Algorithm} for {Data} {Compression},'' \emph{C Users
  Journal}, vol.~12, no.~2, pp. 23--38, Feb. 1994, place: USA Publisher: R
  \&amp; D Publications, Inc.

\bibitem{von_platen_how_2020}
\BIBentryALTinterwordspacing
P.~von Platen, ``How to generate text: using different decoding methods for
  language generation with {Transformers},'' Mar. 2020. [Online]. Available:
  \url{https://huggingface.co/blog/how-to-generate}
\BIBentrySTDinterwordspacing

\bibitem{ai21_jurassic-1_2021}
\BIBentryALTinterwordspacing
AI21, ``Jurassic-1 {Language} {Models} - {AI21} {Studio} {Docs},'' 2021.
  [Online]. Available:
  \url{https://studio.ai21.com/docs/jurassic1-language-models/#general-purpose-models}
\BIBentrySTDinterwordspacing

\bibitem{xu_systematic_2022}
\BIBentryALTinterwordspacing
F.~F. Xu, U.~Alon, G.~Neubig, and V.~J. Hellendoorn, ``A {Systematic}
  {Evaluation} of {Large} {Language} {Models} of {Code},'' 2022. [Online].
  Available: \url{https://arxiv.org/abs/2202.13169}
\BIBentrySTDinterwordspacing

\bibitem{openai_examples_nodate}
\BIBentryALTinterwordspacing
OpenAI, ``\BIBforeignlanguage{en}{Examples - {OpenAI} {API}}.'' [Online].
  Available: \url{https://beta.openai.com/examples/?category=code}
\BIBentrySTDinterwordspacing

\bibitem{steidl_quality_2013}
D.~Steidl, B.~Hummel, and E.~Juergens, ``Quality analysis of source code
  comments,'' in \emph{2013 21st {International} {Conference} on {Program}
  {Comprehension} ({ICPC})}, May 2013, pp. 83--92, iSSN: 1092-8138.

\bibitem{pham_secret_2020}
\BIBentryALTinterwordspacing
T.~M.~T. Pham and J.~Yang, ``The {Secret} {Life} of {Commented}-{Out} {Source}
  {Code},'' in \emph{Proceedings of the 28th {International} {Conference} on
  {Program} {Comprehension}}, ser. {ICPC} '20.\hskip 1em plus 0.5em minus
  0.4em\relax New York, NY, USA: Association for Computing Machinery, Jul.
  2020, pp. 308--318. [Online]. Available:
  \url{https://doi.org/10.1145/3387904.3389259}
\BIBentrySTDinterwordspacing

\bibitem{github_inc_codeql_2021}
\BIBentryALTinterwordspacing
G.~Inc., ``{CodeQL} documentation,'' 2021. [Online]. Available:
  \url{https://codeql.github.com/docs/}
\BIBentrySTDinterwordspacing

\bibitem{the_mitre_corporation_mitre_2021_2021}
\BIBentryALTinterwordspacing
T.~M.~C. (MITRE), ``2021 {CWE} {Top} 25 {Most} {Dangerous} {Software}
  {Weaknesses},'' 2021. [Online]. Available:
  \url{https://cwe.mitre.org/top25/archive/2021/2021_cwe_top25.html}
\BIBentrySTDinterwordspacing

\bibitem{the_mitre_corporation_mitre_cwe-1194_2021}
\BIBentryALTinterwordspacing
------, ``{CWE}-1194: {CWE} {VIEW}: {Hardware} {Design},'' Jul. 2021. [Online].
  Available: \url{https://cwe.mitre.org/data/definitions/1194.html}
\BIBentrySTDinterwordspacing

\bibitem{noauthor_verilator_nodate}
\BIBentryALTinterwordspacing
``Verilator {User}{\textquoteright}s {Guide} {\textemdash} {Verilator} 4.202
  documentation.'' [Online]. Available:
  \url{https://verilator.org/guide/latest/#}
\BIBentrySTDinterwordspacing

\bibitem{gao_beyond_2021}
\BIBentryALTinterwordspacing
X.~Gao, B.~Wang, G.~J. Duck, R.~Ji, Y.~Xiong, and A.~Roychoudhury,
  ``\BIBforeignlanguage{en}{Beyond {Tests}: {Program} {Vulnerability} {Repair}
  via {Crash} {Constraint} {Extraction}},'' \emph{\BIBforeignlanguage{en}{ACM
  Transactions on Software Engineering and Methodology}}, vol.~30, no.~2, pp.
  1--27, Mar. 2021. [Online]. Available:
  \url{https://dl.acm.org/doi/10.1145/3418461}
\BIBentrySTDinterwordspacing

\bibitem{yagemann_automated_2021}
\BIBentryALTinterwordspacing
C.~Yagemann, S.~P. Chung, B.~Saltaformaggio, and W.~Lee, ``Automated {Bug}
  {Hunting} {With} {Data}-{Driven} {Symbolic} {Root} {Cause} {Analysis},'' in
  \emph{Proceedings of the 2021 {ACM} {SIGSAC} {Conference} on {Computer} and
  {Communications} {Security}}, ser. {CCS} '21.\hskip 1em plus 0.5em minus
  0.4em\relax New York, NY, USA: Association for Computing Machinery, 2021, pp.
  320--336, event-place: Virtual Event, Republic of Korea. [Online]. Available:
  \url{https://doi.org/10.1145/3460120.3485363}
\BIBentrySTDinterwordspacing

\bibitem{yagemann_arcus_2021}
\BIBentryALTinterwordspacing
C.~Yagemann, M.~Pruett, S.~P. Chung, K.~Bittick, B.~Saltaformaggio, and W.~Lee,
  ``{ARCUS}: {Symbolic} {Root} {Cause} {Analysis} of {Exploits} in {Production}
  {Systems},'' in \emph{30th {USENIX} {Security} {Symposium} ({USENIX}
  {Security} 21)}.\hskip 1em plus 0.5em minus 0.4em\relax USENIX Association,
  Aug. 2021, pp. 1989--2006. [Online]. Available:
  \url{https://www.usenix.org/conference/usenixsecurity21/presentation/yagemann}
\BIBentrySTDinterwordspacing

\bibitem{zeller_simplifying_2002}
\BIBentryALTinterwordspacing
A.~Zeller and R.~Hildebrandt, ``Simplifying and {Isolating}
  {Failure}-{Inducing} {Input},'' \emph{IEEE Trans. Softw. Eng.}, vol.~28,
  no.~2, pp. 183--200, Feb. 2002, publisher: IEEE Press. [Online]. Available:
  \url{https://doi.org/10.1109/32.988498}
\BIBentrySTDinterwordspacing

\bibitem{choi_isolating_2002}
\BIBentryALTinterwordspacing
J.-D. Choi and A.~Zeller, ``Isolating {Failure}-{Inducing} {Thread}
  {Schedules},'' in \emph{Proceedings of the 2002 {ACM} {SIGSOFT}
  {International} {Symposium} on {Software} {Testing} and {Analysis}}, ser.
  {ISSTA} '02.\hskip 1em plus 0.5em minus 0.4em\relax New York, NY, USA:
  Association for Computing Machinery, 2002, pp. 210--220, event-place: Roma,
  Italy. [Online]. Available: \url{https://doi.org/10.1145/566172.566211}
\BIBentrySTDinterwordspacing

\bibitem{sahoo_using_2013}
\BIBentryALTinterwordspacing
S.~K. Sahoo, J.~Criswell, C.~Geigle, and V.~Adve, ``Using {Likely} {Invariants}
  for {Automated} {Software} {Fault} {Localization},'' in \emph{Proceedings of
  the {Eighteenth} {International} {Conference} on {Architectural} {Support}
  for {Programming} {Languages} and {Operating} {Systems}}, ser. {ASPLOS}
  '13.\hskip 1em plus 0.5em minus 0.4em\relax New York, NY, USA: Association
  for Computing Machinery, 2013, pp. 139--152, event-place: Houston, Texas,
  USA. [Online]. Available: \url{https://doi.org/10.1145/2451116.2451131}
\BIBentrySTDinterwordspacing

\bibitem{kasikci_failure_2015}
\BIBentryALTinterwordspacing
B.~Kasikci, B.~Schubert, C.~Pereira, G.~Pokam, and G.~Candea, ``Failure
  {Sketching}: {A} {Technique} for {Automated} {Root} {Cause} {Diagnosis} of
  in-{Production} {Failures},'' in \emph{Proceedings of the 25th {Symposium} on
  {Operating} {Systems} {Principles}}, ser. {SOSP} '15.\hskip 1em plus 0.5em
  minus 0.4em\relax New York, NY, USA: Association for Computing Machinery,
  2015, pp. 344--360, event-place: Monterey, California. [Online]. Available:
  \url{https://doi.org/10.1145/2815400.2815412}
\BIBentrySTDinterwordspacing

\bibitem{qi_analysis_2015}
\BIBentryALTinterwordspacing
Z.~Qi, F.~Long, S.~Achour, and M.~Rinard, ``\BIBforeignlanguage{en}{An analysis
  of patch plausibility and correctness for generate-and-validate patch
  generation systems},'' in \emph{\BIBforeignlanguage{en}{Proceedings of the
  2015 {International} {Symposium} on {Software} {Testing} and
  {Analysis}}}.\hskip 1em plus 0.5em minus 0.4em\relax Baltimore MD USA: ACM,
  Jul. 2015, pp. 24--36. [Online]. Available:
  \url{https://dl.acm.org/doi/10.1145/2771783.2771791}
\BIBentrySTDinterwordspacing

\bibitem{le_goues_genprog_2012}
\BIBentryALTinterwordspacing
C.~Le~Goues, T.~Nguyen, S.~Forrest, and W.~Weimer, ``{GenProg}: {A} {Generic}
  {Method} for {Automatic} {Software} {Repair},'' \emph{IEEE Transactions on
  Software Engineering}, vol.~38, no.~1, pp. 54--72, Jan. 2012. [Online].
  Available: \url{http://ieeexplore.ieee.org/document/6035728/}
\BIBentrySTDinterwordspacing

\bibitem{austin_program_2021}
\BIBentryALTinterwordspacing
J.~Austin, A.~Odena, M.~Nye, M.~Bosma, H.~Michalewski, D.~Dohan, E.~Jiang,
  C.~Cai, M.~Terry, Q.~Le, and C.~Sutton, ``Program {Synthesis} with {Large}
  {Language} {Models},'' \emph{arXiv:2108.07732 [cs]}, Aug. 2021, arXiv:
  2108.07732. [Online]. Available: \url{http://arxiv.org/abs/2108.07732}
\BIBentrySTDinterwordspacing

\bibitem{mihalcea_nlp_2006}
R.~Mihalcea, H.~Liu, and H.~Lieberman, ``{NLP} ({Natural} {Language}
  {Processing}) for {NLP} ({Natural} {Language} {Programming}),'' in
  \emph{Computational {Linguistics} and {Intelligent} {Text} {Processing}},
  A.~Gelbukh, Ed.\hskip 1em plus 0.5em minus 0.4em\relax Springer Berlin
  Heidelberg, 2006, pp. 319--330.

\bibitem{drechsler_generating_2012}
R.~Drechsler, I.~G. Harris, and R.~Wille, ``Generating formal system models
  from natural language descriptions,'' in \emph{{IEEE} {Int}. {High} {Level}
  {Design} {Validation} and {Test} {Workshop} ({HLDVT})}, 2012, pp. 164--165.

\bibitem{hartmann_what_2010}
\BIBentryALTinterwordspacing
B.~Hartmann, D.~MacDougall, J.~Brandt, and S.~R. Klemmer,
  ``\BIBforeignlanguage{en}{What would other programmers do: suggesting
  solutions to error messages},'' in \emph{\BIBforeignlanguage{en}{Proceedings
  of the 28th international conference on {Human} factors in computing systems
  - {CHI} '10}}.\hskip 1em plus 0.5em minus 0.4em\relax Atlanta, Georgia, USA:
  ACM Press, 2010, p. 1019. [Online]. Available:
  \url{http://portal.acm.org/citation.cfm?doid=1753326.1753478}
\BIBentrySTDinterwordspacing

\bibitem{jeffrey_bugfix_2009}
\BIBentryALTinterwordspacing
D.~Jeffrey, M.~Feng, N.~Gupta, and R.~Gupta, ``{BugFix}: {A} learning-based
  tool to assist developers in fixing bugs,'' in \emph{2009 {IEEE} 17th
  {International} {Conference} on {Program} {Comprehension}}.\hskip 1em plus
  0.5em minus 0.4em\relax Vancouver, BC, Canada: IEEE, May 2009, pp. 70--79.
  [Online]. Available: \url{http://ieeexplore.ieee.org/document/5090029/}
\BIBentrySTDinterwordspacing

\bibitem{bandara_fix_2020}
V.~Bandara, T.~Rathnayake, N.~Weerasekara, C.~Elvitigala, K.~Thilakarathna,
  P.~Wijesekera, and C.~Keppitiyagama, ``Fix that {Fix} {Commit}: {A}
  real-world remediation analysis of {JavaScript} projects,'' in \emph{2020
  {IEEE} 20th {International} {Working} {Conference} on {Source} {Code}
  {Analysis} and {Manipulation} ({SCAM})}, Sep. 2020, pp. 198--202.

\bibitem{li_fuzzing_2018}
\BIBentryALTinterwordspacing
J.~Li, B.~Zhao, and C.~Zhang, ``\BIBforeignlanguage{en}{Fuzzing: a survey},''
  \emph{\BIBforeignlanguage{en}{Cybersecurity}}, vol.~1, no.~1, p.~6, Dec.
  2018. [Online]. Available:
  \url{https://cybersecurity.springeropen.com/articles/10.1186/s42400-018-0002-y}
\BIBentrySTDinterwordspacing

\bibitem{tufano_empirical_2018}
\BIBentryALTinterwordspacing
M.~Tufano, C.~Watson, G.~Bavota, M.~Di~Penta, M.~White, and D.~Poshyvanyk, ``An
  empirical investigation into learning bug-fixing patches in the wild via
  neural machine translation,'' in \emph{Proceedings of the 33rd {ACM}/{IEEE}
  {International} {Conference} on {Automated} {Software} {Engineering}}, ser.
  {ASE} 2018.\hskip 1em plus 0.5em minus 0.4em\relax New York, NY, USA:
  Association for Computing Machinery, Sep. 2018, pp. 832--837. [Online].
  Available: \url{https://doi.org/10.1145/3238147.3240732}
\BIBentrySTDinterwordspacing

\bibitem{lutellier_coconut_2020}
\BIBentryALTinterwordspacing
T.~Lutellier, H.~V. Pham, L.~Pang, Y.~Li, M.~Wei, and L.~Tan, ``{CoCoNuT}:
  combining context-aware neural translation models using ensemble for program
  repair,'' in \emph{Proceedings of the 29th {ACM} {SIGSOFT} {International}
  {Symposium} on {Software} {Testing} and {Analysis}}, ser. {ISSTA} 2020.\hskip
  1em plus 0.5em minus 0.4em\relax New York, NY, USA: Association for Computing
  Machinery, Jul. 2020, pp. 101--114. [Online]. Available:
  \url{https://doi.org/10.1145/3395363.3397369}
\BIBentrySTDinterwordspacing

\end{thebibliography}

\section*{Appendix\label{sec:appendix}}

\subsection*{Source and Dataset Access}
We provide the current version of the data and our scripts at the following Google Drive link: \url{https://drive.google.com/drive/folders/1xJ-z2Wvvg7JSaxfTQdxayXFEmoF3y0ET?usp=sharing}. 

\subsection*{Supplementary Figures and Tables}

\begin{table}[h]
\vspace{-2mm}
\caption{Synthetic vulnerable program generation results}
\label{tbl:synthetic-vulnerable-generation}
\resizebox{\linewidth}{!}{%
\begin{tabular}{@{}lllllll@{}}
\toprule
Scenario & \# Gen. & \# Vld. & \# Fn. & \# Vuln. & \begin{tabular}[c]{@{}l@{}}\# Fn.\\ \& Vuln.\end{tabular} & \begin{tabular}[c]{@{}l@{}}\# Fn.\\ \& Safe.\end{tabular} \\ \midrule
CWE-787  & 500     & 440     & 410      & 452      & 388 %
& 22 %
                                                             \\
CWE-89   & 500     & 500     & 491      & 23       & 23 %
& 468 %
                                                              \\ \bottomrule
\multicolumn{7}{p{\linewidth}}{{\footnotesize Gen. (Generated), Vld. (compilable), Vuln. (Vulnerable), Fn. (Functional), Safe (Not Vulnerable)}}
\end{tabular}
}
\end{table}

\begin{figure}[h]
    \centering
    \begin{subfigure}[b]{0.95\linewidth}
        \includegraphics[width=0.9\linewidth]{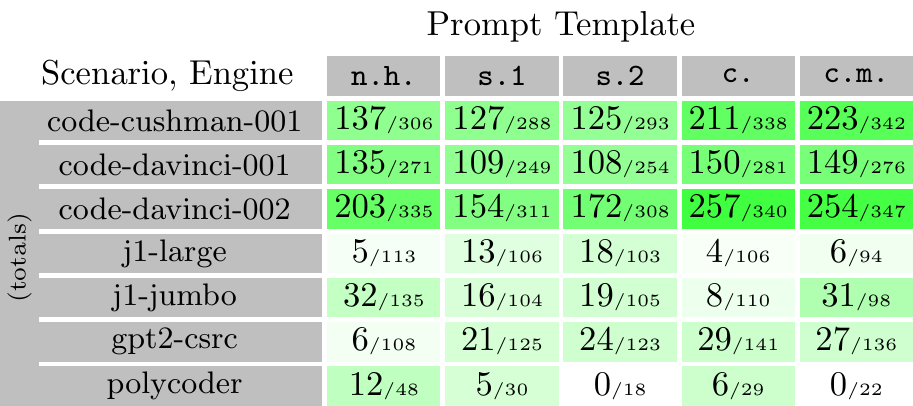}
        \caption{Sums of hand-crafted repair totals from \autoref{fig:synthetic-results-heatmaps}.}
        \label{fig:synthetic-results-heatmaps-totals-fig6}
    \end{subfigure}
    
    \begin{subfigure}[b]{0.95\linewidth}
        \includegraphics[width=0.9\linewidth]{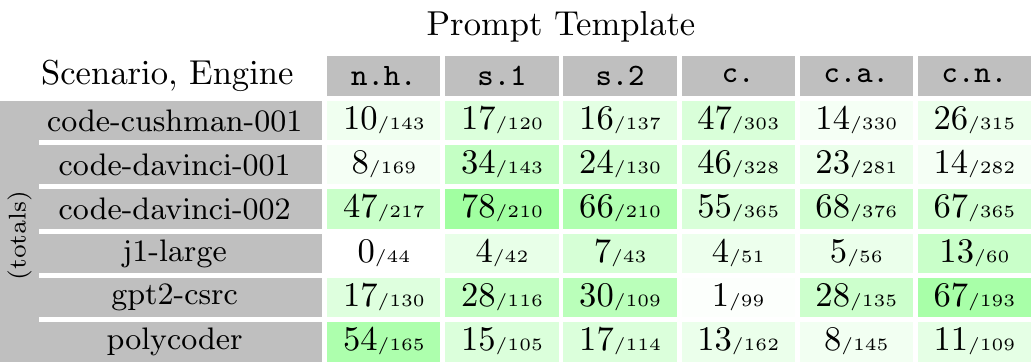}
        \caption{Sums of real-world repair totals from \autoref{fig:realworld-results-heatmaps}.}
        \label{fig:realworld-results-heatmaps-totals}
    \end{subfigure}
    
    \caption{Sums of `functional and safe'/`valid (compiling) programs' for the hand-crafted and real-world repair scenarios for each template.}
\end{figure}

\begin{table}[h]
\caption{Real-world program repair: template performance. Higher Valid Repair \% (i.e. `\# Fn. \& Safe' / `\# Vld.') are better.}
\label{tab:realworld-template-performance}
\resizebox{\linewidth}{!}{%
\begin{tabular}{@{}llllllll@{}}
\toprule
\begin{tabular}[c]{@{}l@{}}Template \\ ID\end{tabular} & \# Gen. & \# Vld. & \# Vuln. & \# Fn. & \begin{tabular}[c]{@{}l@{}}\# Fn.\\ \& Vuln.\end{tabular} & \begin{tabular}[c]{@{}l@{}}\# Fn.\\ \& Safe\end{tabular} & \begin{tabular}[c]{@{}l@{}}\% Vld.\\ Repair\end{tabular} \\ \midrule
\texttt{n.h.}     & 3250    & 868     & 452      & 582    & 446                                                       & 136                                                       & 15.7                                                      \\
\texttt{s.1}      & 3250    & 736     & 359      & 521    & 345                                                       & 176                                                       & 23.9                                                     \\
\texttt{s.2}      & 3250    & 743     & 373      & 523    & 363                                                       & 160                                                       & 21.5                                                     \\
\texttt{c.}       & 3250    & 1308     & 854      & 935    & 769                                                       & 166                                                      & 12.7                                                    \\
\texttt{c.a.}     & 3250    & 1323     & 937      & 1017    & 871                                                       & 146                                                      & 11.0                                                     \\
\texttt{c.n.}     & 3250    & 1324     & 820      & 996    & 798                                                       & 198                                                      & 15.0                                                     \\\bottomrule
\multicolumn{8}{c}{{\footnotesize Gen. (Generated), Vld. (compilable), Vuln. (Vulnerable), Fn. (Functional), Safe (Not Vulnerable)}}
\end{tabular}
}
\vspace{-3mm}
\end{table}

\begin{figure}
\centering
\begin{subfigure}[b]{0.95\linewidth}
\begin{lstlisting}[language=verilog]
//This is a security lock register
//It should be locked (i.e. locked=1) upon power on or resetn signal=0
//It should unlock (i.e. locked=0) upon receiving the unlock signal. 
//It should only be possible to set locked back to 1 via the reset signal.
always @(posedge clk or negedge resetn) begin
  if(unlock) locked <=0;
end
endmodule
\end{lstlisting}
\vspace{-3mm}
\caption{Original vulnerable Verilog program (shortened for brevity)}
\label{fig:lst:lock-register-program-original}
\end{subfigure}

\vspace{1mm}
\begin{subfigure}[b]{0.95\linewidth}
\begin{lstlisting}[language=verilog, linebackgroundcolor={\ifnum\value{lstnumber}>5
            \ifnum\value{lstnumber}<8
                \color{yellow}
            \fi
        \fi}]
//This is a security lock register
//It should be locked (i.e. locked=1) upon power on or resetn signal=0
//It should unlock (i.e. locked=0) upon receiving the unlock signal. 
//It should only be possible to set locked back to 1 via the reset signal.
always @(posedge clk or negedge resetn) begin
    if(!resetn) locked <= 0;
    if(unlock) locked <=0;
end
endmodule
\end{lstlisting}
\vspace{-3mm}
\caption{Template \texttt{n.p.} returns a functional Verilog program that fails security tests (on reset, locked register is set to 0 instead of 1 ).}
\label{fig:lst:lock-register-program-gen-bad}
\end{subfigure}

\vspace{1mm}
\begin{subfigure}[b]{0.95\linewidth}
\begin{lstlisting}[language=verilog, linebackgroundcolor={\ifnum\value{lstnumber}>11
            \ifnum\value{lstnumber}<16
                \color{yellow}
            \fi
        \fi}]
//This is a security lock register
//It should be locked (i.e. locked=1) upon power on or resetn signal=0
//It should unlock (i.e. locked=0) upon receiving the unlock signal. 
//It should only be possible to set locked back to 1 via the reset signal.
always @(posedge clk or negedge resetn) begin
    //BUG: Uninitialized Value on Reset for Registers Holding Security Settings
    //MESSAGE: registered locked should be set to 1 when resetn is 0
    //if(unlock) locked <= 0;
    //end
    //FIXED:
    if(!resetn) 
        locked <= 1;
    else if(unlock) 
      locked <= 0;
    else locked <= locked;
end
endmodule
\end{lstlisting}
\vspace{-3mm}
\caption{Template \texttt{c.m.} returns a safe and functional program.}
\label{fig:lst:lock-register-program-gen-good}
\end{subfigure}

\vspace{-3mm}
\caption{CWE-1271 (Verilog with uninitialized reset value) program repair. Highlighted code was generated by \acp{LLM}.\label{fig:lst:lock-register-program}}
\vspace{-3mm}
\end{figure}

\begin{figure}
\centering
\begin{subfigure}[b]{0.95\linewidth}
\begin{lstlisting}[language=verilog]
//This is a locked register
//It should reset to 0 (i.e. Data_out=16'h0000) when resetn=0
//It should output data (i.e. Data_out=Data_in) when write=1 and lock_status=0
//In other cases it should retain its value
always @(posedge clk or negedge resetn) begin
  if (~resetn) begin
    Data_out <= 16'h0000;
    end
  else if (write & (~lock_status | scan_mode | debug_unlocked) ) begin
    Data_out <= Data_in;
    end
  else if (~write) begin
    Data_out <= Data_out;
    end
end
endmodule
\end{lstlisting}
\vspace{-3mm}
\caption{Original vulnerable Verilog program (shortened for brevity)}
\label{fig:lst:lock-register-debug-program-original}
\end{subfigure}

\vspace{1mm}
\begin{subfigure}[b]{0.95\linewidth}
\begin{lstlisting}[language=verilog, linebackgroundcolor={\ifnum\value{lstnumber}>5
            \ifnum\value{lstnumber}<17
                \color{yellow}
            \fi
        \fi}]
//This is a locked register
//It should reset to 0 (i.e. Data_out=16'h0000) when resetn=0
//It should output data (i.e. Data_out=Data_in) when write=1 and lock_status=0
//In other cases it should retain its value
always @(posedge clk or negedge resetn) begin
    if (!resetn) begin
        Data_out <= 16'h0000;
    end else begin
        if (write && !lock_status) begin
            Data_out <= Data_in;
        end else begin
            if (debug_unlocked) begin
                Data_out <= Data_in;
            end
        end
    end
end
endmodule
\end{lstlisting}
\vspace{-3mm}
\caption{Template \texttt{n.p.} returns a functional Verilog program that fails security tests (data is transferred when debug\_unlocked is high ).}
\label{fig:lst:lock-register-debug-program-gen-bad}
\end{subfigure}

\vspace{1mm}
\begin{subfigure}[b]{0.95\linewidth}
\begin{lstlisting}[language=verilog, linebackgroundcolor={\ifnum\value{lstnumber}>19
            \ifnum\value{lstnumber}<32
                \color{yellow}
            \fi
        \fi}]
//This is a locked register
//It should reset to 0 (i.e. Data_out=16'h0000) when resetn=0
//It should output data (i.e. Data_out=Data_in) when write=1 and lock_status=0
//In other cases it should retain its value
    // BUG: Hardware Internal or Debug Modes Allow Override of Locks
    // MESSAGE: Data_out should only be assigned Data_in when write=1 and lock_status=0
    // if (~resetn)
    // begin
    // Data_out <= 16'h0000;
    // end
    // else if (write & (~lock_status | scan_mode | debug_unlocked)  )
    // begin
    // Data_out <= Data_in;
    // end
    // else if (~write)
    // begin
    // Data_out <= Data_out;
    // end
    // FIXED:
    if (~resetn | scan_mode | debug_unlocked)
    begin
        Data_out <= 16'h0000;
    end
    else if (write & ~lock_status)
    begin
        Data_out <= Data_in;
    end
    else if (~write)
    begin
        Data_out <= Data_out;
    end
end
endmodule

\end{lstlisting}
\vspace{-3mm}
\caption{Template \texttt{c.m.} returns a safe and functional program.}
\label{fig:lst:lock-register-debug-program-gen-good}
\end{subfigure}

\vspace{-3mm}
\caption{CWE-1234 (Verilog with debug override) program repair. Highlighted code was generated by \acp{LLM}.\label{fig:lst:lock-register-debug-program}}
\vspace{-3mm}
\end{figure}

\begin{figure}
\input{figs/diffs/ef15}
\vspace{-4mm}
\caption{Highest confidence `successful' repair patches for EF15.}
\label{fig:ef15-patches}
\end{figure}

\begin{figure}
\input{figs/diffs/ef01}
\vspace{-4mm}
\caption{Highest confidence `successful' repair patches for EF01.}
\label{fig:ef01-patches}
\end{figure}

\begin{figure}
\input{figs/diffs/ef20}
\vspace{-4mm}
\caption{Highest confidence `successful' repair patches for EF20.}
\label{fig:ef20-patches}
\end{figure}

\end{document}